\documentclass[a4paper,11pt,notitlepage]{article}

\usepackage{amsmath}
\usepackage{latexsym}
\usepackage{amssymb}
\usepackage{amsfonts}
\usepackage{verbatim}

\newcommand{\C}{ \mathbb{C} }
\newcommand{\R}{ \mathbb{R} }

\newcommand{\tr}{\mathop{\mathrm{Tr}}}
\newcommand{\rank}{\mathop{\mathrm{rank}}}
\newcommand{\sgn}{\mathop{\mathrm{sgn}}}

\newcommand{\diag}{\mathop{\mathrm{diag}}}
\newcommand{\re}{\mathop{\mathrm{Re}}}
\newcommand{\im}{\mathop{\mathrm{Im}}}

\renewcommand{\vec}[1]{\boldsymbol{#1}}

\author{Yan V. Fyodorov$^{(a)(b)}$ and Boris A. Khoruzhenko$^{(c)}$\\
$^{(a)}$ Institut f\"{u}r Theoretische Physik,\\ Universit\"{a}t
zu
        K\"{o}ln, 50937 K\"{o}ln, Germany\\
        $^{(b)}$School of Mathematical Sciences, University of Nottingham,\\ Nottingham, NG7 2RD, U.K. \\
        $^{(c)}$School of Mathematical Sciences, Queen Mary,
        University of London,\\ London E1 4NS, U.K.}

\title{ A few remarks on Colour-Flavour Transformations,
truncations of random unitary matrices, Berezin reproducing
kernels and Selberg type integrals}

\date{19 October 2006}

\begin{document}

\maketitle

\begin{abstract}
We investigate diverse relations of the colour-flavour
transformations (CFT) introduced by Zirnbauer in \cite{Z1,Z2} to
various topics in random matrix theory and multivariate analysis,
such as measures on truncations of unitary random matrices, Jacobi
ensembles of random matrices, Berezin reproducing kernels and a
generalization of the Selberg integral due to Kaneko, Kadell and
Yan involving the Schur functions. Apart from suggesting explicit
formulas for bosonic CFT for the unitary group in the range of
parameters beyond that in \cite{Z2}, we also suggest an
alternative variant of the transformation, with integration going
over an unbounded domain of a pair of Hermitian matrices. The
latter makes possible evaluation of certain averages in random
matrix theory.

\vspace{3ex}

Short title: A few remarks on Colour-Flavour Transformations

\vspace{3ex}


\end{abstract}

\section{Introduction}
\label{section1}

Let $(\vec{x}_{j})_{j=1}^m$ and $(\vec{y}_{j})_{j=1}^m$ be two sets
of column-vectors in $\C^N$. By convention, the column-vectors are
regarded as matrices consisting of a single column, and  we shall
use the star to denote the Hermitian conjugate (complex conjugate
transpose) of a matrix. Let $U(N)$ stand for the group of unitary
matrices of size $N\times N$ equipped with the Haar measure
$d\mu_H(U)$ fixed by the normalization condition
$\int_{U(N)}d\mu_H(U)=1$. Further, consider the matrix ball $Q^{*}Q
\le I_m$ in $\C^{m\times m}$, the space of complex $m\times m $
matrices, equipped with the unit mass measure
\begin{equation}\label{mub}
d\mu^{B}_{N,m} (Q) = \hbox{const.}\, \det (I_m-Q^*Q)^{N-2m}\ (dQ),
\qquad N\ge 2m,
\end{equation}
where $I_m$ is identity matrix and $(dQ)$ is the cartesian volume
element in $\C^{m\times m}$,
\[
(dQ)= \prod_{j,k=1}^md \re Q_{jk} d\im Q_{jk}.
\]
With these notations in mind, the following remarkable identity,
called the bosonic Colour-Flavour Transformation (bCFT),
\begin{equation}\label{bcft}
\int_{U(N)}e^{\sum_{j=1}^m (\vec{y}^{*}_{j}U \vec{x}_{j} +
\vec{x}^{*}_{j}U^{*} \vec{y}_{j}) }\ d\mu_H(U) = \int_{Q^{*}Q\le
I_m} e^{\sum_{j,k=1}^m (Q_{jk}\vec{x}^{*}_{k}\vec{x}_{j} +
 (Q^*)_{jk} \vec{y}^{*}_{k} \vec{y}_{j} ) }\ d\mu^{B}_{N,m}
(Q),
\end{equation}
is known to hold for $N\ge 2m$. Similarly, the fermionic
Colour-Flavour transformation (fCFT) asserts that
\begin{equation}\label{fcft}
\int_{U(N)}e^{\sum_{j=1}^m (\vec{\chi}^{*}_{j}U \vec{\psi}_{j} +
\vec{\psi}^{*}_{j}U^{*} \vec{\chi}_{j}) }\ d\mu_H(U) =
\int_{\C^{m\times m}} e^{\sum_{j,k=1}^m
(Q_{jk}\vec{\chi}^{*}_{k}\vec{\chi}_{j} -
 (Q^*)_{jk} \vec{\psi}^{*}_{k} \vec{\psi}_{j} ) }\ d\mu^{F}_{N,m}
(Q),
\end{equation}
where now $\vec{\chi}_{j}$, $\vec{\psi}_{j}$, $\vec{\chi}^*_{j}$ and
$\vec{\psi}^*_{j}$ are vectors with anti-commuting components. The
$\vec{\chi}_{j}$ and $\vec{\psi}_{j}$ are column-vectors and the
$\vec{\chi}^*_{j}$ and $\vec{\psi}^*_{j}$ are row-vectors. In
contrast to (\ref{bcft}), there is no restriction on $m$ and $N$ in
(\ref{fcft}). The integration on the right-hand side in (\ref{fcft})
is over the entire space of complex $m\times m$ matrices and
\begin{equation}\label{muf}
d\mu^{F}_{N,m} (Q) = \hbox{const.}\, {\det}^{-N-2m}(I_m+Q^{*}Q)\
(dQ).
\end{equation}

Both identities, together with the unifying supersymmetric variant
of Colour-Flavour Transformation (CFT) and extensions to other
classical groups, were originally discovered by Zirnbauer in 1996
\cite{Z1,Z2} and proved by a skillful use of the machinery of
generalized coherent states \cite{Perelomov}. Following Zirnbauer's
approach, variants of the CFT were also obtained for the special
unitary group in \cite{Budczies,SW3,WW}. Later on it was realized,
again by Zirnbauer, see e.g \cite{Zirntrento,ZirnRMT}, that actually
all forms of the CFT are just manifestations of a very deep
algebraic fact related to the so-called Howe duality \cite{Howe}.
Since their introduction, the Colour-Flavour Transformations proved
to be a very useful tool, finding diverse applications in such areas
of physics as lattice gauge theory \cite{Z2,Budczies,SW}, random
network models \cite{AltSim,Zirnhall}, quantum chaos models
\cite{AltZirn,AltGnutz}, and the random matrix theory \cite{Z1,FK}.

In this paper we revisit the simplest case of the unitary group. Our
inspiration comes from noticing a certain similarity between a few
results known in the random matrix theory and the bosonic and
fermionic versions of the CFT. The central role in our investigation
is played by certain integrals involving the so-called Schur
functions $s_{\lambda} (A)$. The latter are explicitly defined for
any $m\times m$ matrix $A$ in terms of its eigenvalues
$a_1,\ldots,a_m$ as
\begin{equation}\label{schur}
s_{\lambda} (A) = s_{\lambda} (a_1, \ldots, a_m)= \frac{\det
\big(a_i^{m+\lambda_j-j}\big)_{i,j=1}^m}{\det\big(a_i^{m-j}\big)_{i,j=1}^m},
\hspace{3ex} \lambda_1\ge \lambda_2 \ge \ldots \ge \lambda_m\ge 0,
\end{equation}
with $\lambda$ being a partition, i.e. a non-increasing sequence of
non-negative integers $\lambda_j$. The Schur functions
$s_{\lambda}(A)$ are symmetric polynomials in the eigenvalues of
$A=(A_{ij})$, and are also polynomials in $A_{ij}$. A concise
introduction to the theory of symmetric functions can be found in
the first chapter of \cite{Macdonald}, see also \cite{Bump}.

The usefulness of Schur functions for our purposes can be traced
to the fact that they are characters of irreducible
representations of the general linear group and its unitary
subgroup and, as a consequence, possess certain properties of
orthogonality, see, in particular, equations (\ref{or2}) and
(\ref{or1}). This makes Schur function expansions a powerful tool
for evaluating integrals over unitary groups as has been already
demonstrated in \cite{Kaz,Balantekin,SW,WW,O,OS,HO}. In this
paper, we use the Schur function expansion technique to extend the
bosonic version of CFT (\ref{bcft}) to the range $N\le 2m\le 2N$.
We also reveal the relation of bCFT and fCFT to several important
matrix integrals due to Berezin, and also to certain
generalizations of the famous Selberg integral due to Kaneko,
Kadell and Yan. Finally, we also derive a new variant of the
bosonic CFT, replacing the integration on the right-hand side in
(\ref{bcft}) with one going over an unbounded matrix domain
parameterized by a pair of Hermitian matrices. Such a
representation is useful for studying regularized inverse spectral
determinants of complex random matrices, the subject of our
earlier work \cite{FK}. The latter work has in fact quite a few
points of intersection with some of the topics discussed in the
present paper.

The fact that the Schur functions play central role in our way of
understanding and extending both bCFT and fCFT  can be traced back
to the above mentioned Howe duality, although we do not exploit the
latter explicitly in the present paper. Without going into any
detailed discussion, we would like to mention that in one of its
incarnations Howe duality can be looked at, see e.g. \cite{BG} or
Chapter 43 in book \cite{Bump}, as the ultimate reason for the
validity of the so-called Cauchy identities
\begin{eqnarray} \label{cauchyid1}
\prod_{i=1}^m\prod_{j=1}^n\frac{1}{1-t_ix_j}
&=&\sum_{\lambda}s_{\lambda}(t_1,\ldots,t_m)s_{\lambda}(x_1,\ldots,x_n)\\
\label{cauchyid2} \prod_{i=1}^m\prod_{j=1}^n
(1+t_ix_j)&=&\sum_{\lambda}s_{\lambda}(t_1,\ldots,t_m)s_{\lambda'}(x_1,\ldots,x_n),
\end{eqnarray}
where $\lambda'$ stands for the conjugate partition. Being ``a point
of very direct connection between representation theory and
combinatorics'' \cite{Bump}, it is therefore no surprise that both
identities (\ref{cauchyid1}) and (\ref{cauchyid2}) play an important
role in obtaining the results of this paper.

The organization of the paper is as follows. In the next section we
will provide, for the reader's convenience, an overview and short
discussion of the most important results following from our approach
to CFT. The subsequent sections will be devoted to technical
derivations of the main formulas.

\section{An overview of the main results.}
\label{section2}

 Our first observation relates the bCFT to truncations of
random unitary matrices. Truncations of random unitary matrices
emerged recently in a different context of quantum chaotic
scattering. Various matrix distributions arising from such
truncations were the subject of a few recent works, see
\cite{ZS,FSom,SM,For1}.

The relation between the bCFT and truncations of unitary matrices
becomes more apparent if one writes (\ref{bcft}) using matrix
notation in the exponential. Introducing two $N\times m$ matrices
$X$ and $Y$ with columns $\vec{x}_1, \ldots, \vec{x}_m$ and
$\vec{y}_1, \ldots, \vec{y}_m$ respectively, one can write
\[
\sum_{j=1}^m (\vec{y}^{*}_{j}U \vec{x}_{j} + \vec{x}^{*}_{j}U^{*}
\vec{y}_{j}) = \tr (Y^*UX + X^*U^{*}Y)=\tr (XY^*U + U^{*}YX^*)
\]
and
\[
X^*X=(\vec{x}_j^*\vec{x}_k)_{j,k=1}^m \hspace{2ex} \hbox{and}
\hspace{2ex} Y^*Y=(\vec{y}_j^*\vec{y}_k)_{j,k=1}^m.
\]
Then the bCFT (\ref{bcft}) takes the following form
\begin{equation}\label{bcft-xy1}
\int_{U(N)}e^{\tr (XY^*U + U^{*}YX^*)}\ d\mu_H(U) =
\int_{Q^{*}Q\le I_m} e^{\tr (X^*X Q + Q^* Y^*Y )}\ d\mu^{B}_{N,m}
(Q).
\end{equation}
Note that the left-hand side in (\ref{bcft-xy1}) is well defined
for any $m$ whilst the integration measure on the right-hand side
is singular if $2m>N$ and care must be taken if one wants to
interpret the above formula for $2m>N$. We argue in Section
\ref{section3} that for $m\le N$ one can rewrite the right-hand
side in a form free from singularities. Indeed, the matrix $XY^*$
and its Hermitian conjugate $Y^*X$ have rank $m\le N$, or, to be
more precise, at most $m$. In view of the invariance of the Haar
measure, this means that the integral on the left-hand side in
(\ref{bcft-xy1}) goes effectively over the principal (top left)
$m\times m$ sub-block $Q$ of the unitary matrix $U$. With this
observation in hand, a straightforward application of the Schur
function expansion for the exponential function $\exp \tr M$
allows one to recast the bCFT in the following form
\begin{equation}\label{b1}
\int_{U(N)}e^{\tr (XY^*U + U^{*}YX^*)}\ d\mu_H(U) =
\int_{Q^{*}Q\le I_m} e^{\tr (X^*X Q + Q^* Y^*Y )}\
d\rho_{N,m\times m} (Q),
\end{equation}
where $d\rho_{N,m\times m} (Q)$ is the image of the Haar measure
under the truncation $U\mapsto Q$. It is known from \cite{FSom} and
\cite{For1}, see also  \cite{Ner1},  that $d\rho_{N,m\times m}
(Q)\propto \det (I_m-Q^*Q)^{N-2m} (dQ)$ for $N\ge 2m$. Thus, as one
would expect, (\ref{b1}) reverts back to the original version of the
bCFT in the interval $N\ge 2m$.

Identity (\ref{b1}) which holds for any $m\le N$ (note that
$d\rho_{N,m\times m} (Q)$ is a unit mass measure and is free from
singularities) provides the basis for our extension of the bCFT to
the range $m\le N$. The boundary case $m=N$ is straightforward.
Indeed, if $m=N$ then no truncation is involved and (\ref{b1}) takes
the form
\[
\int_{U(N)}e^{\tr (Y^*UX + X^*U^{*}Y) }d\mu_H(U)= \int_{U(N)}
e^{\tr (X^*X U + U^{*}Y^*Y) } d\mu_H(U).
\]
The intermediate range $N<2m<2N$ requires additional calculations
which are of interest on their own. Though in this range the measure
$d\rho_{N,m\times m} (Q)$ can still be found explicitly, the
resulting expression is too complicated to be practically used.
However, for our purposes we only need to know the radial part of
$d\rho_{N,m\times m} (Q)$ and this latter can be found in the
explicit form and then used to rewrite the right-hand side of
(\ref{b1}).

As we discuss in Section \ref{section3}, in the range $N<2m<2N$ the
measure $d\rho_{N,m\times m} (Q)$ is supported by the set $\Upsilon$
on the boundary of the matrix ball $Q^*Q\le I_m$ defined by the
condition that $\rank (I_m-Q^*Q)=N-m$.  This set can be parametrized
by the matrices
\begin{equation}\label{quzv}
Q_{UZV^*} = U \left(
  \begin{array}{cc}
    Z & 0 \\
    0 & I_{2m-N} \\
  \end{array}
\right) V^*
\end{equation}
where $Z$ runs through the matrix ball $Z^*Z < I_{N-m}$ in
$\C^{(N-m)\times (N-m)}$, the space of complex $(N-m)\times (N-m)$
matrices, and $U$ and $V$ run through the unitary group $U(m)$.
However, this parametrization is not one-to-one: different
$Q_{UZV^*}$ may represent the same point $Q$ in $\Upsilon$. In the
terminology of \cite{Hua} (page 118) the set of matrices $Q_{UZV^*}$
is a covering of $\Upsilon$. Exploiting such a parametrization, we
obtain the following variant of the bCFT in the range $N<2m<2N$
\begin{eqnarray} \label{cft3}
\lefteqn{\int_{U(N)}e^{\tr (Y^*UX + X^*U^{*}Y) } d\mu_H(U) =} \\
\nonumber && \int_{U(m)} \int_{U(m)} \int_{Z^*Z\le I_{N-m}}
\hspace{-3.5ex} e^{\tr (X^*XQ_{UZV^*}+ Q_{UZV^*}^*Y^*Y)}\
d\mu^{B}_{N, N-m} (Z) d\mu_H(U)d\mu_H(V),
\end{eqnarray}
where $Q_{UZV^*}$ is as defined in (\ref{quzv}) and $d\mu^{B}_{N,
N-m} (Z)$ is the unit mass measure on the matrix ball $Z^*Z\le
I_{N-m}$ defined in (\ref{mub}),
\begin{equation}\label{mub1}
d\mu^{B}_{N, N-m} (Z) = \hbox{const.} \det (I_{N-m} - Z^*Z)^{2m-N}
(dZ), \qquad N \le 2m<2N.
\end{equation}

It has to be mentioned that another variant of extension of the bCFT
to the range $m\le N$,
\begin{equation}\label{w-w}
\int_{U(N)}e^{\tr (Y^*UX + X^*U^{*}Y) } d\mu_H(U) = \int_{U(m)} \det
(VY^*X)^{m-N} e^{\tr (X^*XV^* + V Y^*Y) } d\mu_H(V),
\end{equation}
was obtained recently in \cite{WW}. Clearly, apart from the boundary
case of $N=m$ this formula is different from ours. Interestingly, as
was observed in \cite{WW}, in the range $m < N$, the integral on the
left-hand side in (\ref{w-w}) does not change if the integration
over the unitary group is replaced by the integration over the
special unitary group:
\begin{equation}\label{yw}
\int_{U(N)}e^{\tr (Y^*UX + X^*U^{*}Y) } d\mu_H(U)=
\int_{SU(N)}e^{\tr (Y^*UX + X^*U^{*}Y) } d\mu_H(U), \qquad m<N.
\end{equation}
This means that for $m<N$ our formula (\ref{cft3}) holds without
further changes if one replaces $U(N)$ by $SU(N)$ in the integral on
the left-hand side.

The original CFT (\ref{bcft}) and its extension (\ref{b1}) almost
hide the fact that both integrals, the one on the left-hand side and
the one on the right hand side, depend only on the eigenvalues of
the matrix product $X^*XY^*Y$. In fact, the bosonic CFT can be
stated in the following, slightly more abstract form. If $A$ and $B$
are two $N\times N$ matrices of rank $m\le N$ then
\begin{equation}\label{39}
\int_{U(N)}e^{\tr (AU+U^*B)}d\mu_H(U)= \int_{Q^*Q\le I_m} e^{\tr
(CQ+Q^*D)} \ d\rho_{N,m\times m}(Q)
\end{equation}
for any pair of $m\times m$ matrices $C$ and $D$ such that the
eigenvalues of $CD$  coincide with the non-zero eigenvalues of
$AB$. In the range $N<2m<2N$ the integration over matrices $Q$ on
the right-hand side can be replaced by integration over the
matrices $Q_{UZV^*}$ as in (\ref{cft3}).

Our next observation is that the CFT is related to several
interesting matrix integrals.

It will be apparent from our calculations in Section \ref{section3}
that the bosonic CFT (\ref{b1}) implies the following identity for
the Schur functions
\begin{equation}\label{7a}
\int_{Q^*Q\le I_m} s_{\lambda} (Q^*Q)\  d\rho_{N,m\times
m}(Q)=\frac{s_{\lambda}^2(I_m)}{s_{\lambda}(I_N)}, \qquad m\le N,
\end{equation}
and vice versa, (\ref{7a}) implies (\ref{b1}). In fact, (\ref{7a})
is a particular case (corresponding to $m=n$) of a more general
relation
\begin{equation}\label{sint3}
\int_{Q^*Q\le I_m} s_{\lambda} (Q^*Q)\ d\rho_{N,n\times m}(Q) =
\frac{s_{\lambda}(I_m)s_{\lambda}(I_n)}{s_{\lambda}(I_N)}, \qquad
m,n\le N.
\end{equation}
Here the matrices $Q$ are $n\times m$ and $d\rho_{N,n\times m}(Q)$
is the image of the Haar measure under the truncation of unitary
matrix to its principal $n\times m $ sub-block, $U\mapsto Q$. This
relation and its generalization
\begin{equation}\label{sint3z}
\int_{Q^*Q\le I_m} s_{\lambda} (XQYQ^*)\ d\rho_{N,n\times m}(Q) =
\frac{s_{\lambda}(X)s_{\lambda}(Y)}{s_{\lambda}(I_N)},  \qquad
m,n\le N,
\end{equation}
are simple corollaries of the invariance of the measure
$d\rho_{N,n\times m}(Q)$ with respect to the right and left
multiplication by unitary matrices. Another corollary of this
invariance (and of the orthogonality of the Schur functions, see
(\ref{or1})) are the following orthogonality relations
\begin{equation}\label{sint3y}
\int_{Q^*Q\le I_m} s_{\lambda} (LQ)\overline{s_{\mu}(MQ)}\
d\rho_{N,n\times m} (Q)= \delta_{\lambda,\mu}\
\frac{s_{\lambda}(M^*L)}{s_{\lambda}(I_N)}, , \qquad m,n\le N,
\end{equation}
which hold for arbitrary $m\times n$ matrices $L$ and $M$.
Identities (\ref{sint3z}) and (\ref{sint3y}) are derived in
Section \ref{section3}.

We show that the integration formulas (\ref{sint3})--(\ref{sint3y})
imply several non-trivial matrix integrals, some of which we believe
to be new. In particular, by making use of the Schur function
expansion (\ref{cauchyid1}), we derive the following identity
\begin{eqnarray}\label{jacob}
\lefteqn{\int_{U(N)}\frac{d\mu_H(U)}{\det(I_N-AU)^m\det(I_N-U^*B^*)^n}=}\\
\nonumber & & \hbox{const.} \int_{Z^*Z\le I_{\min(m,n)}}
\frac{1}{\det(I
 -Z^*Z \otimes B^*A)}\ \det
(Z^*Z)^{|n-m|}\det (I-Z^*Z)^{N-m-n} \ (dZ)
\end{eqnarray}
which reduces the group integral on the left to an average over the
Jacobi ensemble of random matrices $Z$ of size $k\times k$,
$k=\min(m,n)$. Identity (\ref{jacob}) holds for $N\ge n+m$ and
$A^*A<I_N$, $B^*B<I_N$ and generalizes our earlier result \cite{FK}
from $n=m$ to the case $n\ne m$. A similar identity holds in the
range $N<m+n<2N$.

In \cite{FK} we obtained  an identity which is dual to (\ref{7a}),
\begin{equation}\label{14}
\int_{\C^{m\times m}} s_{\lambda} (Z^*Z)\
d\mu^{F}_{N,m}(Z)=\frac{s_{\lambda}^2(I_m)}{s_{\lambda^{\prime}}(I_N)}.
\end{equation}
Here $d\mu^{F}_{N,m}$, see (\ref{muf}), is the measure which appears
on the right-hand side in the fermionic version of the CFT
(\ref{fcft}). By making use of the Schur function expansion
(\ref{cauchyid2}), identity (\ref{14}) is equivalent to the matrix
integral
\begin{equation}\label{15}
\int_{U(N)} \det(I_N+AU)^m\det(I_N+U^*B^*)^m\
d\mu_H(U)=\int_{\C^{m\times m}}\hspace{-1ex} \det(I + Z^*Z\otimes
B^*A)\ d\mu^{F}_{N,m} (Z)
\end{equation}
which is dual to the $m=n$ version of (\ref{jacob}). The emergence
of $d\mu^{F}_{N,m} (Z)$ on the right-hand side in (\ref{15}) is not
coincidental. In fact, we show in Section \ref{section4} that the
fermionic version of the CFT (\ref{fcft}) implies (\ref{15})
directly. We expect that identity (\ref{14}) (or, equivalently, the
matrix integral (\ref{15})) should in turn imply the fermionic CFT,
by analogy to the relation between the bosonic CFT and identity
(\ref{7a}). Unfortunately, we have succeeded in verifying such
equivalence only for the simplest case $m=1$.

In the same way as identity (\ref{7a}) allows for an extension to
rectangular matrices, identity (\ref{14}) allows for a similar
extension:
\begin{equation}\label{14a}
\int_{\C^{n\times m}} s_{\lambda} (Q^*Q)\ d\mu^{F}_{N, n\times
m}(Q)= \frac{s_{\lambda}
(I_n)s_{\lambda}(I_m)}{s_{\lambda^{\prime}}(I_N)},
\end{equation}
where $d\mu^{F}_{N, n\times m}(Q)$ is the unit mass measure
\[
d\mu^{F}_{N, n\times m}(Q) = \hbox{const.}\, \det
(I_m+Q^*Q)^{-N-m-n} \ (dQ)
\]
on the space of complex $n\times m$ matrices. Identity (\ref{14a})
holds for any positive integers $N$, $m$ and $n$, and in turn,
implies the identity dual to (\ref{jacob}):
\begin{eqnarray}\label{jacobdual}
\lefteqn{\int_{U(N)} \det(I_N+AU)^m\det(I_N+U^*B^*)^n\ d\mu_H(U)
=}
\\ \nonumber & & \hbox{const.} \int_{\C^{\min(m,n)\times \min(m,n)}}\det(I
 +Z^*Z \otimes B^*A)\ \frac{\det
(Z^*Z)^{|n-m|} }{\det (I+Z^*Z)^{N+m+n}}\ (dZ) ,
\end{eqnarray}
again reducing evaluation of the integral over the unitary group on
the left to evaluation of an integral over a Jacobi ensemble of
random matrices. This is a generalization of our earlier result from
\cite{FK}.

Identities (\ref{7a}) and (\ref{14}) yield another pair of matrix
integrals, again by the way of the Schur function expansions
(\ref{cauchyid1})--(\ref{cauchyid2}):
\begin{equation}\label{19}
\int_{Q^*Q\le I_m} \frac{d\rho_{N,n\times
m}(Q)}{\det(I_n-Z_1Q^*)^N\det(I_n-QZ_2^*)^N}=\frac{1}{\det(I_n-Z_1Z_2^*)^N},
\end{equation}
where $Z_i^*Z_i < I_m$,  and
\begin{equation}\label{20}
\int_{\C^{n\times m}} \det(I_m+Z_1Q^*)^N\det(I_m+QZ_2^*)^N\
d\mu_{N,n\times m}^{F}(Q)=\det(I_m+Z_1Z_2^*)^N.
\end{equation}
These matrix integrals are not new. They are a variant of
integrals obtained by Berezin in his work on quantization in
complex symmetric spaces \cite{B1,B2}.  Berezin proved (\ref{20})
for integer $N$ and (\ref{20}) for a range of real $N$ that
includes $N\ge n+m$. In Section \ref{section4} we discuss this
link and quote some of Berezin's results.

Since identities (\ref{sint3}) and (\ref{14a}) are so useful in the
context of Schur function expansions, we think it is worth to have a
closer look at them. Without loss of generality we can assume that
$n\ge m$. In the bosonic case we also assume that $N\ge n+m$, so
that the integration measure $d\rho_{N,n\times m}(Q)$ in
(\ref{sint3}) is replaced by $\hbox{const.} \det (I_m-Q^*Q)^{N-m-n}
(dQ)$. In the range $N<n+m < 2N$ one can obtain slightly different
formulas by using parametrization (\ref{quzv}), see especially the
integration formula (\ref{sint5}).

The integration in (\ref{sint3}) and (\ref{14a}) is effectively over
the eigenvalues of $Q^*Q$. By making the corresponding change of
variables (see, \cite{Hua} or \cite{Forbook}) one brings the matrix
integral in (\ref{sint3}) to
\begin{equation}\label{30}
\frac{1}{c^N_{n,m}} \int_0^1 \hspace{-1ex}\cdots \int_0^1
s_{\lambda}(x_1, \ldots, x_m) \prod_{j=1}^m\
 x_j^{n-m} (1-x_j)^{N-m-n}\hspace{-1ex}   \prod_{1\le i<j\le m}\hspace{-1.5ex}(x_i-x_j)^2 \
\prod_{j=1}^m dx_j
=\frac{s_{\lambda}(1_n)s_{\lambda}(1_m)}{s_{\lambda}(1_N)}
\end{equation}
and the one in (\ref{14a}) to
\begin{equation}\label{31}
\frac{1}{k^N_{n,m}}\int_0^{\infty} \hspace{-1.5ex} \cdots
\int_0^{\infty}  s_{\lambda}(x_1, \ldots, x_m) \prod_{j=1}^m\
\frac{x_j^{n-m}}{ (1+x_j)^{N+m+n}} \prod_{1\le i<j\le
m}\hspace{-1.5ex}(x_i-x_j)^2 \ \prod_{j=1}^m dx_j=
\frac{s_{\lambda}(1_n)s_{\lambda}(1_m)}{s_{\lambda^{\prime}}(1_N)}.
\end{equation}
The normalization constants in these formulas can be computed with
the help of the celebrated Selberg integral, and are given in
(\ref{32}) and (\ref{33}).

Interestingly, the integral in (\ref{30}) is the $\gamma =1$ case of
the following extension of the Selberg integral due to Kaneko
\cite{Ka} (see also related works by  Kadell \cite{Kad} and Yan
\cite{Yan})
\begin{eqnarray}\label{34}
\lefteqn{ \int_0^1 \hspace{-1ex}\cdots \int_0^1
J^{\frac{1}{\gamma}}_{\lambda}(x_1, \ldots, x_m) \prod_{j=1}^m\
x_j^{p-1} (1-x_j)^{q-1}\hspace{-1ex} \prod_{1\le i<j\le
m}\hspace{-1.5ex}|x_i-x_j|^{2\gamma} \ \prod_{j=1}^m dx_j}\\
\nonumber &=& J^{\frac{1}{\gamma}}_{\lambda}(1_m)\prod_{i=1}^m
\frac{\Gamma (i\gamma +1)\Gamma (\lambda_i+p+\gamma (m-i))\Gamma
(q+\gamma (m-i))}{\Gamma (1+\gamma)\Gamma (\lambda_i +p+q+\gamma
(2m-i-1)) }.
\end{eqnarray}
Here $J^{\frac{1}{\gamma}}_{\lambda}(x)$ are the Jack symmetric
functions \cite{Stanley}, $J^{\frac{1}{\gamma}}_{\lambda}(x)$ are
proportional to the Schur functions if $\gamma =1$,
$J^{1}_{\lambda}(x)=H_{\lambda}s_{\lambda}(x)$, where $H_{\lambda}$
is a coefficient independent of $x$. Similarly, (\ref{31}) is the
$\gamma =1$  case of the integral
\begin{equation}\label{35}
\int_0^{\infty} \hspace{-1.5ex} \cdots \int_0^{\infty}
J^{\frac{1}{\gamma}}_{\lambda}(t_1, \ldots, t_m) \prod_{j=1}^m\
\frac{t_j^{p-1}}{(1+t_j)^{p+q+2(m-1)\gamma}} \hspace{-1ex}
\prod_{1\le i<j\le m}\hspace{-1.5ex}|t_i-t_j|^{2\gamma} \
\prod_{j=1}^m dt_j.
\end{equation}
To the best of our knowledge, the latter integral is not evaluated
yet for $\gamma \not=1$ and we consider its computation as an
interesting open problem. In particular, we note that although for
the zero partition $\lambda_1=\lambda_2 =\ldots =\lambda_m=0$ the
substitution $x=1/(1-t)$ reduces the integral in (\ref{35}) to the
one in (\ref{34}), this substitution does not preserve the Jack
symmetric functions, and hence the integral in (\ref{35}) for
non-zero $\lambda$ requires a separate evaluation.

Surprisingly, the $\gamma=1$ case is simple in the sense that the
integral in (\ref{35}) (and the one in (\ref{34})) can be
evaluated by elementary means, as we demonstrate in the end of
Section \ref{section4}.

Section \ref{section5} of our paper is devoted to yet another
alternative ("deformed" ) version of the bosonic CFT with a
different integration manifold in the integral on the right-hand
side of (\ref{bcft}). Consider the manifold of matrices $(Q_1, Q_2)$
in $\C^{m\times m}\times \C^{m\times m} $ parametrized as follows
\begin{equation}\label{q1q2}
Q_1=TPT^*, \qquad Q_2=(T^*)^{-1}PT^{-1},
\end{equation}
where $T$ runs through the general linear group, $T\in GL_m(\C)$,
and $P$ runs through the set of diagonal matrices
\begin{equation}\label{q1q2a}
P=\diag (p_1, \ldots, p_m); \hspace{1ex} -1 \le p_{j} \le 1,
\hspace{1ex}  j=1, \ldots , m,
\end{equation}
and \emph{introduce} the integration ``measure''
\begin{equation}\label{q1q2b}
(dQ_1dQ_2)=\prod_{1\le j<k\le m}(p_j^2-p_k^2)^2 \prod_{j=1}^m
p_jdp_j\ d\mu_H(T),
\end{equation}
where $ d\mu_H(T)=(dT)/\det (T^*T)^m$ is the invariant measure on
$GL_m(\C)$. Note that $p_j$ change sign, and therefore $(dQ_1dQ_2)$
is not a proper positive measure, see \cite{F2} for a discussion.
Then we claim the validity of the following variant of the bosonic
CFT ($N\ge 2m$)
\begin{equation}\label{cft5}
\int_{U(N)}e^{-i\tr (Y^*UX + X^*U^{*}Y)}d\mu_H(U)=\int \det
(I_m-Q_1Q_2)^{N-2m} e^{-i\tr (Q_1Y^*Y+Q_2X^*X)}(dQ_1dQ_2).
\end{equation}
Note the most important features of this formula are: (i)
integration domain in (\ref{cft5}) is unbounded, in contrast to the
bounded domain $Q^*Q\le I_m$ as in the standard bCFT
(\ref{bcft-xy1}) and (ii) the matrices $Q_1$ and $Q_2$ are
Hermitian. We finish that section by demonstrating that the last
property makes the new version of bCFT indispensable when evaluating
expectations values of negative powers of certain (regularized)
spectral determinants.

\section{bCFT from truncations of random unitary matrices}
\label{section3}

In this section we first address the problem of evaluation of
integrals of the form
\begin{equation}\label{sint}
\int_{Q^*Q\le I_m} f(Q^*Q)\ d\rho_{N,n\times m} (Q),
\end{equation}
where $d\rho_{N,n\times m} (Q)$ is the image of the Haar measure
under the truncation of unitary matrices as defined below. Then we
apply the obtained formulas to derive the bCFT in the form of
equations (\ref{b1}) and (\ref{cft3}).

Let $U$ be an $N\times N$ unitary matrix and $m$ and $n$ be
positive integer numbers, $m\le n\le N$. Partition $U$ into the
four blocks
\begin{equation}\label{block}
U= \left(
  \begin{array}{cc}
    Q & R \\
    P & S \\
  \end{array}
\right)
\end{equation}
where the top left block $Q$ is $n\times m$. Partition (\ref{block})
defines the map $\omega: U \to Q$ from the unitary group into the
matrix ball $Q^*Q\le I_m$ in $\C^{n\times m}$, the space of complex
$n\times m$ matrices. Under this map the Haar measure $d\mu_H (U)$
on $U(N)$ induces a measure on the matrix ball $Q^*Q\le I_m$ which
we shall denote by $d\rho_{N,n\times m}(Q)$. The unitarity of $U$
imposes constraints on its sub-blocks. In particular,
\begin{equation}\label{qp}
Q^*Q+P^*P =I_m.
\end{equation}
If $N\ge m+n$ then, generically, the matrix $P^*P$ has rank $m$,
and the image of $U(N)$ under the map $\omega$ is the entire
matrix ball $Q^*Q\le I_m$. In this case the measure
$d\rho_{N,n\times m}(Q)$ has been previously computed in
\cite{Ner1,FSom} for square matrices $Q$ and in \cite{For1} for
rectangular matrices,
\begin{equation}\label{2mleN}
d\rho_{N,n\times m} (Q) = \hbox{const.}\det (I_m-Q^{*}Q)^{N-m-n}
(dQ), \qquad N\ge n+m,
\end{equation}
where $(dQ)$ is the cartesian volume element in $\C^{n\times m}$.
By making an appropriate change of variables, see e.g.
\cite{Forbook} or one of the calculations below, the integral in
(\ref{sint}) can then be reduced to the familiar form
\begin{equation}\label{2mleN-1}
\int_{Z^*Z\le I_m}  f(Z^*Z) \det (Z^*Z)^{n-m} \det
(I_m-Z^{*}Z)^{N-m-n} (dZ)
\end{equation}
of an average over the so-called Jacobi ensemble  \cite{Forbook}
of random $m\times m$ matrices $Z$ and can be evaluated by the
standard tools of the random matrix theory.

If $N<m+n$, i.e. $N-n< m$,  then, generically, the matrix $P^*P$
in (\ref{qp}) has rank $N-n$   and, as a consequence, $z=1$ is an
eigenvalue of $Q^*Q$ of multiplicity $m+n-N$. Therefore, in this
case the image of $U(N)$ under the map $\omega$ is a submanifold
of the boundary of the matrix ball $Q^*Q\le I_m$. This submanifold
is defined by the equations
\begin{equation}\label{qman}
\left. \frac{d^k}{dz^k} \det(zI_m-Q^*Q)\right|_{z=1} =0,
\hspace{3ex} \hbox{$k=0,1, \ldots , m+n-N-1$.}
\end{equation}
An expression for $d\rho_{N,n\times m}$ in this case can be obtained
by following the method of calculation of $d\rho_{N,n\times m}$ for
$N\ge n+m$ that was suggested in \cite{FSom} and extended in
\cite{For1}. It makes use of the matrix integral \cite{F1}
\[
\int \frac{e^{i\tr FX} (dF) }{{\det}^l (F-zI_k)}= c_{l,k}
 e^{iz\tr X} \det X^{l-k}, \hspace{2ex} \hbox{$l\ge k$, $\im z >0$ and $X>0$}
\]
over Hermitian $k\times k$ matrices $F$. Extension of this
calculation to the case of $N<n+m$ requires evaluation of the
above integral for positive semi-definite matrices $X$ of rank
$l<k$. Such evaluation was given in \cite{JN}. Using this result
one can calculate $d\rho_{N,n\times m}$ for $N<n+m$, however, the
resulting expression contains many delta-functional factors and is
not very useful for direct applications. Fortunately, evaluation
of integrals (\ref{sint}) is a simpler task and can be
accomplished with the help of the following standard calculation
from multivariate analysis.

Consider $\C^{l\times k}$, the space of complex $l\times k$
matrices. If $l \ge k $ then any matrix $P\in \C^{l\times k}$ of
rank $k$ can be uniquely written as $P=HT$ where $T=(T_{ij})
\in\C^{k\times k}$ is upper-triangular with positive diagonal
elements and $H\in \C^{l\times k}$ is such that $H^*H=I_k$.
Correspondingly, the cartesian volume element $(dP)$ in $\C^{l\times
k}$ transforms as follows, see e.g. \cite{Muirhead,Mathai},
\begin{equation}\label{dP-1}
(dP) = \hbox{const} \times     \det (TT^*)^{l-k} \prod_{i=1}^k
T_{ii}^{2(k-i)+1} (dT)(H^*dH)
\end{equation}
where
\[
(dT) = \prod_{1\le i \le k} T_{ii}\prod_{1\le i < j \le k}
  d\re T_{ij} d\im T_{ij}
\]
and $(H^*dH)$ is the invariant volume element\footnote{For details
of its construction, see \cite{James,Muirhead}} on the Stiefel
manifold $V_k(\C^l)\cong U(l)/U(l-k)$ of complex $l\times k$
matrices with orthonormal columns. It follows from (\ref{dP-1}) that
if $f$ is a function on $\C^{k\times k}$ then
\begin{equation}\label{ir1}
\int_{\C^{l\times k}} f(P^*P) (dP) = \hbox{const.} \int_{\C^{k\times
k}} f(Z^*Z) \det (Z^*Z)^{l-k}
 (dZ),
\end{equation}
where we have used (\ref{dP-1}) twice, at first making the
substitution $P=HT$ and then making the reverse substitution
$T=V^{-1}Z$ with $Z\in \C^{k\times k}$ and $V\in U(k)$. Similarly,
if $g$ is a function on $\C^{l\times l}$ then
\begin{equation}\label{ir2}
\int_{\C^{l\times k}} g(PP^*) (dP) = \hbox{const.}
\int_{V_k(\C^l)}\int_{\C^{k\times k}} g(HZZ^*H^*) \det (ZZ^*)^{l-k}
 (dZ) (H^*dH).
\end{equation}
The integration rules (\ref{ir1}) and (\ref{ir2}) hold for $l\ge k$.
If $l<k$ then
\begin{equation}\label{ir3}
\int_{\C^{l\times k}} f(P^*P) (dP) = \hbox{const.}
\int_{V_l(\C^k)}\int_{\C^{l\times l}} f(HZ^*ZH^*) \det (Z^*Z)^{k-l}
 (dZ) (H^*dH)
\end{equation}
and
\begin{equation}\label{ir4}
\int_{\C^{l\times k}} g(PP^*) (dP) = \hbox{const.} \int_{\C^{l\times
l}} g(ZZ^*) \det (ZZ^*)^{k-l}
 (dZ).
\end{equation}

Returning to integrals (\ref{sint}), let us consider, alongside with
$\omega$, another map defined by partition (\ref{block}),
\[
\tau: U\to H=\left(\begin{array}{cc} Q\\P
\end{array} \right).
\]
It maps the unitary group $U(N)$ onto the Stiefel manifold of
$N\times m$ matrices $H$ with orthonormal columns, $H^*H=I_m$.
Obviously, the image of the Haar measure $d\mu_H(U)$ under this map
is invariant with respect to the right and left multiplications by
unitary matrices. Since the Stiefel manifold is a coset space of the
unitary group, such invariant measure is unique up to a
multiplicative constant. Therefore,
\begin{equation}\label{inv1}
\int_{U(N)} g(\tau (U)) \ d\mu_H(U) =\int_{U(N)} g(H) \ d\mu_H(U) =
\hbox{const.} \int_{V_m(\C^N)} g(H) (H^*dH).
\end{equation}
Sometimes it is convenient to write the invariant measure on
$V_m(\C^N)$ as a singular measure on $\C^{N\times m}$
\begin{equation}\label{inv2}
\int_{V_m(\C^N)} g(H) (H^*dH) = \hbox{const.} \int_{C^{N\times m}}
g(H) \delta(H^*H-I_m) (dH),
\end{equation}
where $(dH)$ is the cartesian volume element in $\C^{N\times m}$ and
$\delta (H^*H-I_m)$ is the matrix delta function on the space of
Hermitian matrices,
\[
\delta (A)=\prod_{j=1}^m \delta (A_{jj}) \prod_{1\le j<k \le m}
\delta(\re A_{jk}) \delta(\im A_{jk}).
\]

Thinking of the truncation $\omega: U\to Q$, where $U\in U(N)$ and
$Q\in \C^{n\times m}$, as a composition of the two successive
truncations
\[
U\mapsto \left(\begin{array}{cc} Q\\P \end{array} \right) \mapsto Q,
\]
we have, by (\ref{inv1})--(\ref{inv2}),
\begin{equation}\label{42}
\int_{Q^*Q\le I_m} \hspace{-2ex} f(Q^*Q) d\rho_{N,n\times m } (Q) =
\hbox{const.} \int_{\C^{n\times m}} \left( \int_{\C^{(N-n)\times m}}
\delta(Q^*Q+P^*P-I_m)(dP)\right) f(Q^*Q) (dQ).
\end{equation}
If $N\ge n+m$ then, by making use of the integration rule
(\ref{ir1}), one can replace the integration over $Q$ and $P$ in
the integral on the right-hand side by integrations over $m\times
m$ matrices $Z$ and $F$, respectively, thus reducing the integral
to the following one
\[
\int_{\C^{m\times m}}\left( \int_{\C^{m \times m}}
\delta(Z^*Z+F^*F-I_m)\det (F^*F)^{N-n-m}\ (dF)\right) f(Z^*Z) \det
(Z^*Z)^{n-m}\ (dZ)\,.
\]
Performing the integration over $F$ one obtains
\begin{equation}\label{sint6}
\int_{ Q^*Q\le I_m} f(Q^*Q) d\rho_{N,n\times m} (Q) = \hbox{const.}
\int_{Z^*Z\le I_m} f(Z^*Z)\det (I_m-Z^*Z)^{N-n-m}\det (Z^*Z)^{n-m}
(dZ),
\end{equation}
in agreement with (\ref{2mleN-1}).

If $N<n+m$, i.e. $N-n <m$, then by making use of the integration
rule (\ref{ir1}) to replace integration over $Q$ by integration over
$Z$ as above and the integration rule (\ref{ir3}) to replace
integration over $P$ by integration over $FH^*$, where  $H\in
V_{N-n}(\C^m)$ and $F\in \C^{(N-n)\times (N-n)}$, one arrives, after
carrying out the integration over $Z$, at
\begin{eqnarray*}
\lefteqn{\int_{ Q^*Q\le I_m} f(Q^*Q) d\rho_{N,n\times m} (Q)
=\hbox{const.} \times
}\\[1ex] && \int_{V_{N-n}(\C^m)} \int_{F^*F\le I_{N-n}} f(I_m-HF^*FH^*)
\det (F^*F)^{n+m-N} \det (I_{N-n}-F^*F)^{n-m}\ (dF)(H^*dH).
\end{eqnarray*}
If the function $f$ is invariant with respect to the conjugation by
unitary matrices, i.e. $f(UAU^*)=f(A)$ for unitary $U$, then
\[
f(I_m-HF^*FH^*)= f\left(I_m - \left(
                                \begin{array}{cc}
                                  F^*F & 0 \\
                                  0 & 0 \\
                                \end{array}
                              \right)
\right) \hspace{2ex} \hbox{for any $H$ such that $H^*H=I_m$},
\]
and the integral on the right-hand side simplifies,
\begin{eqnarray}\nonumber
\lefteqn{\int_{ Q^*Q\le I_m} f(Q^*Q) d\rho_{N,n\times m} (Q)=
\hbox{const.} \times
} \\[1ex] \label{sint5} &&   \hspace{-2ex}\int_{Z^*Z\le I_{N-n}} \hspace{-1ex}
f\left(
    \begin{array}{cc}
        Z^*Z & 0 \\
        0 & I_{m+n-N} \\
    \end{array}
\right) \det (I_{N-n}-Z^*Z)^{n+m-N} \det (Z^*Z)^{n-m}\ (dZ)
\end{eqnarray}

If the function $f(A)$ is invariant under the conjugation by
unitary matrices then it is effectively a function of the
eigenvalues $a_1, \ldots, a_m$ of $A=(A_{jk})_{j,k=1}^m$. There is
one class of such functions for which the integral in (\ref{sint})
can be easily computed, see (\ref{sint3}). These are the Schur
functions $s_{\lambda}$ (\ref{schur}). The Schur functions are the
characters of irreducible polynomial representations of the
general linear group. Such representations remain irreducible when
restricted to the unitary subgroup of the general linear group.
The orthogonality of matrix elements of irreducible
representations as functions on the unitary group then implies the
following integration formulae, see e.g. \cite{Macdonald}, p.445,
\begin{equation}\label{or2}
\int_{U(m)}s_{\lambda} (AU BU^*)d\mu_H(U) = \frac{s_{\lambda} (A)
s_{\lambda} (B)}{s_{\lambda} (I_m)}
\end{equation}
and
\begin{equation}\label{or1}
\int_{U(m)}s_{\lambda} (AU)\overline{s_{\mu}
(BU)}d\mu_H(U)=\delta_{\lambda,\mu}\frac{s_{\lambda}
(AB^*)}{s_{\lambda} (I_m)}.
\end{equation}
Our calculation of the integral in (\ref{sint}) for $f=s_{\lambda}$
rests on the integration formula (\ref{or2}) and the invariance of
the measure $d\rho_{N,n\times m}$ with respect to the left and right
multiplications by unitary matrices.

Let $X$ be an $n\times n$ matrix and $Y$ be an $m\times m$ matrix.
Consider the integral
\begin{equation}\label{sint1}
\int_{U(N)} s_{\lambda} (AUBU^*)\ d\mu_H(U)= \frac{s_{\lambda} (A)
s_{\lambda} (B)}{s_{\lambda} (I_N)}
\end{equation}
where $A$ and $B$ are the block diagonal matrices $A=\diag (X,0)$,
and $B=\diag (Y,0)$. It is apparent from (\ref{schur}) that
\begin{equation}\label{red}
s_{\lambda} (a_1, a_2, \ldots, a_m, 0, \ldots, 0) =\left\{
\begin{array}{ll}
s_{\lambda} (a_1, a_2, \ldots, a_m) & \hbox{if $l(\lambda)\le m$}\\
0 & \hbox{if $l(\lambda)>m$}
\end{array}
\right.
\end{equation}
where $l(\lambda)$ is the length of $\lambda$ (the number of
non-zero parts $\lambda_j$)\footnote{By convention, one does not
distinguish between partitions which differ merely by the number of
zero parts, i.e. $(\lambda_1, \ldots, \lambda_m)=(\lambda_1, \ldots,
\lambda_m, 0, \ldots, 0)$}. Therefore the integral in (\ref{sint1})
vanishes for $l(\lambda)>m$ (recall that $m\le n$) and is equal to
$s_{\lambda}(X)s_{\lambda}(Y)/s_{\lambda}(I_N)$ if $l(\lambda)\le
m$.

On the other hand, the non-zero eigenvalues of the $N\times N$
matrix $AUBU^*$ coincide with those of the $m\times m$ matrix
$XQYQ^*$ where $Q$ is the principal $n\times m$ block of $U$, see
(\ref{block}). Therefore
\begin{equation}\label{sint2}
\int_{U(N)}s_{\lambda} (AUBU^*)\ d\mu_H(U)=\int_{Q^*Q\le I_m}
s_{\lambda} (XQYQ^*)\ d\rho_{N,n\times m}(Q).
\end{equation}
The measure $d\rho_{N,n\times m}(Q)$ is invariant with respect to
the right multiplication by unitary matrices. Hence
\[
\int_{Q^*Q\le I_m} s_{\lambda} (XQYQ^*)\ d\rho_{N,n\times m}(Q)=
\int_{U(m)} \int_{Q^*Q\le I_m} s_{\lambda} (Q^*XQVYV^*)\
d\rho_{N,n\times m}(Q) d\mu_H(V),
\]
where we have used the invariance of $s_{\lambda}(M_1M_2\ldots)$
under the cyclic permutations of the matrices $M_j$. Reverting the
order of integrations and applying the integration rule (\ref{or2}),
one obtains
\[
\int_{Q^*Q\le I_m} s_{\lambda} (XQYQ^*)\ d\rho_{N,n\times m}(Q)=
\frac{s_{\lambda}(Y)}{s_{\lambda}(I_m)}\ \int_{Q^*Q\le I_m}
\hspace{-2ex}s_{\lambda} (XQQ^*)\ d\rho_{N,n\times m}(Q).
\]
The measure $d\rho_{N,n\times m}(Q)$ is also invariant with respect
to the left multiplications by unitary matrices. Repeating the above
procedure, one decouples $X$ and $QQ^*$ thus obtaining
\[
\int_{Q^*Q\le I_m} s_{\lambda} (XQYQ^*)\ d\rho_{N,n\times m}(Q)=
\frac{s_{\lambda}(X)s_{\lambda}(Y)}{s_{\lambda}(I_m)s_{\lambda}(I_n)}\
\int_{Q^*Q\le I_m} \hspace{-2ex}s_{\lambda} (Q^*Q)\ d\rho_{N,n\times
m}(Q).
\]
On comparing this to (\ref{sint1}) and (\ref{sint2}), one concludes
that
\begin{equation}\label{sint3a}
\int_{Q^*Q\le I_m} s_{\lambda} (XQYQ^*)\ d\rho_{N,n\times m}(Q) =
\frac{s_{\lambda}(X)s_{\lambda}(Y)}{s_{\lambda}(I_N)}
\end{equation}
of which (\ref{sint3}) is a particular case of $X=I_n$, $Y=I_m$. It
is worth mentioning that the quotient
${s_{\lambda}(I_n)s_{\lambda}(I_m)}/{s_{\lambda}(I_N)}$  can be
easily evaluated in terms of $\lambda_j$'s by recalling Weyl's
dimension formula
\begin{equation}\label{eq2:8}
s_{\lambda}(I_n)=\left\{\prod_{1\le i <j\le
m}(\lambda_i-i-\lambda_j+j)\right\}\times \prod_{j=1}^m
\frac{(n+\lambda_j-j)!}{(m+\lambda_j-j)!(n-j)!}
\end{equation}
which holds for any integer $n\ge m\ge l(\lambda)$.

By repeating the argument which was used to evaluate the integral in
(\ref{sint3a}), one can extend the integration formulae (\ref{or2})
and (\ref{or1}) from integrals over unitary group to integrals over
complex matrices provided that the integration measure is invariant
with respect to the left \emph{and} right multiplication by unitary
matrices. For example, if $L$ and $M$ are $m\times n$ matrices and
$d\rho (Q)$ is a measure on $\C^{n\times m}$ invariant with respect
to the right and left multiplication by unitary matrices and such
that the integral below converges absolutely then
\begin{equation}\label{or4}
\int_{\C^{n\times m}} s_{\lambda} (LQ)\overline{s_{\mu}(MQ)}\ d\rho
(Q)= \delta_{\lambda,\mu}\ \frac{s_{\lambda}(M^*L)}{s_{\lambda}(I_n)
s_{\lambda}(I_m)} \ \int_{\C^{n\times m}} s_{\lambda} (Q^*Q) d\rho
(Q).
\end{equation}
In particular, if $d\rho$ is the projection of the Haar measure on
the matrix ball $Q^*Q\le I_m$, i.e., $d\rho=d\rho_{N,n\times m}$,
then (\ref{or4}) and  (\ref{sint3a}) imply the orthogonality
relation (\ref{sint3y}).

We would like to make two remarks at this point. One is that
equations (\ref{sint6}) and (\ref{sint5}) effectively give the joint
probability distribution of the singular values of the truncations
of random unitary matrices. This distribution has an interesting
symmetry: Consider two square truncations, $Q_1$ of size $m\times m$
and $Q_2$ of size $(N-m)\times (N-m)$. Assuming that $2m \le N$,
$x=1$ is the singular value of $Q_2$ of multiplicity $N-2m$ and the
remaining $m$ singular values of $Q_2$ have the same distribution as
the $m$ singular values of $Q_1$.

The other is that some of our calculations can be repeated, almost
verbatim, for truncations of random orthogonal matrices. In
particular, let $Q$ be the top left block of size $n\times m$ of
random orthogonal matrix $O$ of size $N\times N$, $m\le n\le N$.
Denote by $d\rho_{O(N),n\times m}$ the image of the Haar measure
under the map $O\mapsto Q$.  Then, by repeating the steps of the
above derivation of (\ref{sint6}) one obtains the integration
formula
\begin{eqnarray*}
\lefteqn{\int_{ Q^t Q\le I_m} f(Q^t Q) d\rho_{O(N),n\times m} (Q) =
\hbox{const.}}\\[1ex]  & & \int_{X^tX\le I_m} f(X^tX)\det
(I_m-X^tX)^{\frac{1}{2}(N-n-m+1)}\det (X^tX)^{\frac{1}{2}(n-m)}
(dX), \qquad N\ge n+m,
\end{eqnarray*}
where the integration on the right is over real $m\times m$ matrices
$X$. This integration formula is not new and was previously obtained
in \cite{For1} by a different method. If $N < n+m$ and $f(A)$ is a
function on $\R^{m\times m}$ which is invariant under the
conjugation by orthogonal matrices then a similar formula holds
\begin{eqnarray*}
\lefteqn{\int_{ Q^tQ\le I_m} f(Q^*Q) d\rho_{O(N),n\times m} (Q)=
\hbox{const.} \times
} \\[1ex] \label{sint5o} &&   \hspace{-2ex}\int_{X^tX\le I_{N-n}} \hspace{-1ex}
f\left(
  \begin{array}{cc}
    X^*X & 0 \\
    0 & I_{m+n-N} \\
  \end{array}
\right) \det (I_{N-n}-X^*X)^{\frac{1}{2}(n+m-N+1)} \det
(X^tX)^{\frac{1}{2}(n-m)}\ (dX).
\end{eqnarray*}
Formula (\ref{or2}) also has its analogue for orthogonal matrices
\begin{equation*}
\int_{O(m)} C_{\lambda} (XOYO^t)\ d\mu_H(O) =
\frac{C_{\lambda}(X)C_{\lambda}(Y)}{C_{\lambda}(I_m)},
\end{equation*}
where $C_\lambda$ are the so-called zonal polynomials. For the
definition of zonal polynomials and their properties see
\cite{Muirhead}. As a consequence, formula (\ref{sint3a}) also has
its analogue for real matrices: one just replaces Schur functions
in (\ref{sint3a}) by zonal polynomials $C_\lambda$. However, the
orthogonality relations (\ref{or1}) and (\ref{or4}) do not seem to
have analogues for real matrices.

Now we are in a position to derive the bosonic CFT formula in the
range $0\le m\le N$. Our approach is based on the Schur function
expansion for the exponential $e^{\tr M}$  combined with the
orthogonality relation (\ref{or4}). This yields the bCFT in the
form (\ref{b1}). The integration formulas (\ref{sint6}) and
(\ref{sint5}) then lead to the specialization (\ref{bcft}) in the
interval $2m \le N$ and (\ref{cft3}) in  the interval $N<2m<2N$.

Recall that the matrices $X$ and $Y$ are $N\times m$. The singular
value decomposition for $XY^*$ reads
\begin{equation}\label{svd}
XY^*=V \tilde D W^*, \hspace{3ex} \tilde D =\left(
  \begin{array}{cc}
    D & 0 \\
    0 & 0 \\
  \end{array}
\right),
\end{equation}
where $V$ and $W$ are unitary matrices of size $N\times N$ and $D$
is a diagonal matrix of size $m\times m$. The entries of $D$ are
exactly the square roots of the eigenvalues of the matrix
$X^*XY^*Y$, so that
\begin{equation}\label{add3}
s_{\lambda} (D^2) = s_{\lambda} (X^*XY^*Y)
\end{equation}
In view of (\ref{svd}) the left-hand side of (\ref{b1}) takes the
form of an integral over truncations of unitary matrices:
\begin{equation}\label{add1}
\int\limits_{U(N)}\hspace{-1ex}e^{\tr (XY^*U + U^{*}YX^*) }d\mu_H(U)
= \int\limits_{U(N)} \hspace{-1ex}e^{\tr (\tilde D U + U^*\tilde D)
} d\mu_H(U) = \int\limits_{Q^*Q\le I_m}\hspace{-1ex} e^{\tr (DQ +
Q^{*}D) } d\rho_{N,m\times m} (Q),
\end{equation}
where $Q$ is the top left $m\times m$ block of $U$ and
$d\rho_{N,m\times m} (Q)$ is the image of the Haar measure under the
map $U\mapsto Q$.

The exponential function $e^{\tr M}$ is a symmetric function of the
eigenvalues of $M$ and as such can be expanded in Schur functions,
see e.g. \cite{Balantekin}:
\begin{equation}\label{40}
e^{\tr M} = \sum_{\lambda} c_{\lambda} s_{\lambda} (M).
\end{equation}
The sum above is over all partitions $\lambda$ of length
$l(\lambda)\le N$ where $N$ is the matrix dimension of $M$.
However, if the matrix $M$ has only $m\le N$ non-zero eigenvalues
then, in view of (\ref{red}), the sum is effectively over
partitions $\lambda$ of length $l(\lambda)\le m$. The coefficients
$c_{\lambda}$ in (\ref{40}) can be computed in terms of $\lambda$:
\begin{equation}\label{40-1}
c_{\lambda}=\det \left( \frac{1}{(\lambda_j-j+i)!}\right)_{i,j=1}^N
= \frac{ \prod_{1\le i<j\le l}
(\lambda_i-i-\lambda_j+j)}{\prod_{j=1}^l(l+\lambda_j-j)!} ,
\end{equation}
where $l$ is the length of partition $\lambda$ and, by convention,
$\frac{1}{k!}=0$ for negative $k$. Note that $c_{\lambda}$ are
independent of the matrix dimension $N$. In Appendix we give an
analytic method of computing the coefficients of Schur function
expansion for multiplicative functionals of eigenvalues of $M$, to
complement the algebraic method of \cite{Balantekin}.

Let us now expand $e^{\tr DQ}$ and $e^{\tr Q^*D}$  in the integral
on the left in (\ref{add1}) in the Schur functions and apply the
orthogonality relation (\ref{or4}) (recall that $d\rho_{N,m\times
m}(Q)$ is invariant with respect to the right and left
multiplications by unitary matrices). This yields
\begin{equation}\label{1-1}
\int_{Q^*Q\le I_m} e^{\tr (DQ + Q^{*}D) } d\rho_{N,m\times m} (Q)
= \sum_{\lambda}\, c^2_{\lambda}\, \frac{s_{\lambda}
(X^*XY^*Y)}{s_{\lambda}^2 (I_m)}\,
 \int_{Q^*Q\le I_m} s_{\lambda} (Q^*Q) d\rho_{N,m\times m}(Q),
\end{equation}
where we have used (\ref{add3}).  Applying (\ref{or4}) again, now
in the opposite direction, and folding the Schur function
expansions for the exponential functions, one obtains
\begin{equation}\label{40-2}
\sum_{\lambda}\, c^2_{\lambda}\,
\frac{s_{\lambda}(X^*XY^*Y)}{s_{\lambda}^2 (I_m)}\,
 \int_{Q^*Q\le I_m} \hspace{-2ex} s_{\lambda} (Q^*Q) d\rho_{N,m\times m}(Q)=
\int_{Q^*Q\le I_m} \hspace{-2ex} e^{\tr (X^*XQ + Q^{*}Y^*Y) }
d\rho_{N,m\times m} (Q),
\end{equation}
and hence the bosonic CFT for $m\le N$ in the form of equation
(\ref{b1}).

The right-hand side of (\ref{b1}) takes different forms depending
on $m$. If $N\ge 2m$ then $ d\rho_{N,m\times m} (Q) =\hbox{const.}
\det (I_m-Q^*Q)^{N-2m}$, see (\ref{2mleN}), and we are back to
formula (\ref{bcft-xy1}). In the range $N<2m<2N$ the measure
$d\rho_{N,m\times m} (Q)$ is not very handy in its explicit form.
However, by (\ref{40-2}), the integral on the right-hand side in
(\ref{b1}), is effectively written in terms of the integrals
\begin{equation}\label{sint4}
\int_{Q^*Q\le I_m}\, s_{\lambda}(Q^*Q)\, d\rho_{N, m\times m} (Q),
\end{equation}
which makes it possible to evaluate it explicitly by making use of
(\ref{sint4}).

Recall, that in the range $N< 2m < 2N$ the measure $d\rho_{N,m\times
m}$ is supported by the set $\Upsilon$ of $m\times m$ matrices $Q$
such that the rank of $I_m-Q^*Q$ is $N-m$ and $Q^*Q\le I_m$. This
set can be parameterized by matrices $Q_{UZV^*}$, see (\ref{quzv}),
and the integral in (\ref{sint4}) is effectively over matrices $Z$,
see (\ref{sint5}). Such parametrization allows one to rewrite the
right-hand side of (\ref{b1}) in a more explicit form. Indeed, it is
apparent that
\[
s_{\lambda}\left(
  \begin{array}{cc}
    Z^*Z & 0 \\
    0 & I_{2m-N} \\
  \end{array} \right)=s_{\lambda}(Q_{UZV^*}^* Q_{UZV^*}).
\]
Hence, by (\ref{sint5}),
\begin{equation}\label{quzv1}
\int_{Q^*Q\le I_m} \hspace{-2.5ex} s_{\lambda} (Q^*Q)\ d\rho_{N,
m\times m} (Q) = \int_{U(m)^2} \int_{Z^*Z< I_{N-m}}
\hspace{-3.5ex}s_{\lambda}(Q_{UZV^*}^*Q_{UZV^*})\ d\mu^{B}_{N, N-m}
(Z) d\mu_H(U)d\mu_H(V),
\end{equation}
where $d\mu^{B}_{N, N-m} (Z)$ is the measure defined in
(\ref{mub1}). Substituting the obtained expression for integral
(\ref{sint4}) into (\ref{1-1}), one obtains the variant of the CFT
in the range $N<2m<2N$ as presented in (\ref{cft3}).

We would like to finish this section with two observations. Firstly,
instead of reducing the group integral on the left-hand side in
(\ref{b1}) to an integral over truncations of unitary matrices, one
can directly expand the exponentials in the group integral and then
apply integration formulas (\ref{or1}) and (\ref{sint3}), the latter
in the right-to-left direction. On this way one easily obtains the
bCFT in the form of equation (\ref{39}).

Secondly, our derivation of the bCFT does not use the explicit
expression (\ref{40-1}) for the coefficients $c_{\lambda}$ of the
Schur function expansion for the exponential function. All that is
needed of $c_{\lambda}$'s is the property
\begin{equation}\label{add2}
c_{(\lambda_1, \ldots, \lambda_m, 0, \ldots, 0)}= c_{(\lambda_1,
\ldots, \lambda_m)}.
\end{equation}
Therefore our calculation will go through for any convergent series
$g(A)=\sum_{\lambda}c_{\lambda}S_{\lambda}(A)$ provided that the
coefficients $c_{\lambda}$ satisfy (\ref{add2}) yielding  the
following generalization of the bCFT formula (\ref{b1})
\begin{equation}\label{bcftgen}
\int_{U(N)}|g((XY^*U)|^2 d\mu_H(U) = \int_{QQ^{*}\le I_m}  g(X^*X
Q)\overline{g(Y^*Y Q)}\,d\rho_{N,m\times m} (Q).
\end{equation}

\section{CFT, reproducing kernels and Selberg integrals}
\label{section4}

 In this section we investigate relations between the CFTs
and several interesting matrix integrals.

It is apparent from the calculations in the previous section that
the bosonic CFT (\ref{b1}) implies the identity (\ref{7a}) and
vice versa.

Consider now the matrix integral
\begin{equation}\label{10a}
\int_{U(N)}\frac{d\mu_H(U)}{\det(I_N-AU)^m\det(I_N-U^*B^*)^m}=\sum_{\lambda}\
\frac{s_{\lambda}^2(I_m)}{s_{\lambda}(I_N)}\ s_{\lambda} (B^*A).
\end{equation}
where $A^*A<I_N$ and $B^*B<I_N$. One can see that the integral on
the left-hand side coincides with the series on the right-hand
side by recalling the Cauchy identity (\ref{cauchyid1}). It is a
natural generalization of the well-known expansion of the inverse
determinant $1/\det(I-M)$ in terms of the complete symmetric
functions $h_r$ of the eigenvalues $(z_1, \ldots, z_m)$ of $M$,
\[
\frac{1}{\det (I-M)}=\prod_{j=1}^m\frac{1}{1-z_j}=
\sum_{r=0}^{\infty}\ h_r(z_1, \ldots, z_m).
\]
Expanding each of the two determinants in (\ref{10a}) with the
help of (\ref{cauchyid1}) and then applying the orthogonality
relation (\ref{or1}) one obtains the series on the right-hand side
in (\ref{10a}). By making use of (\ref{7a}) and (\ref{cauchyid1}),
one can fold this series back to a matrix integral, now over
matrices $Q$:
\begin{equation}\label{10aa}
\int_{U(N)}\frac{d\mu_H(U)}{\det(I_N-AU)^m\det(I_N-U^*B^*)^m}=
\int_{Q^*Q\le I_m}\frac{d\rho_{N,m\times m}(Q)}{\det(I_{mN}
 -Q^*Q \otimes B^*A)}, \hspace{3ex} m\le N.
\end{equation}
Thus, the bosonic CFT (\ref{b1}) implies (\ref{10aa}) and vice
versa.

In the range $2m \le N$ the measure $d\rho_{N,m\times m}(Q)$
coincides with $d\mu^{B}_{N,m} (Q)$ of (\ref{mub}), and (\ref{10aa})
reads
\[
\int_{U(N)}\frac{d\mu_H(U)}{\det(I_N-AU)^m\det(I_N-U^*B^*)^m}=
\int_{Q^*Q\le I_m}\frac{d\mu^{B}_{N,m} (Q)}{\det(I
 -Q^*Q \otimes B^*A)}.
\]
This identity was obtained in our earlier work \cite{FK}.

In the range $N<2m<2N$ one can replace the integration over matrices
$Q$ by integration over matrices $Q_{UZV^*}$, see (\ref{quzv}) and
(\ref{quzv1}), and (\ref{10aa}) takes this form
\begin{equation*}%\label{12}
 \int\limits_{U(N)}\hspace{-0.5ex}\frac{d\mu_H(U)}{\det(I_N-AU)^m\det(I_N-U^*B^*)^m} =
 \int\limits_{Z^*Z\le I_{N-m}}\hspace{-2ex}\frac{d\mu^{B}_{N,N-m}(Z)}{\det(I
 -Z^*Z \otimes B^*A)\det(I_{N}
 - B^*A)^{2m-N} }.
\end{equation*}
The integration on the right-hand side is over complex $(N-m)\times
(N-m)$ matrices $Z$ and the integration measure is given in
(\ref{mub1}). This identity is new.

If $m=N$ then $d\rho_{N, m\times m}$ is the Haar measure on $U(N)$,
and (\ref{10aa}) reads
\begin{equation*}%\label{13}
\int_{U(N)}\frac{d\mu_H(U)}{\det(I_N-AU)^N\det(I_N-U^*B^*)^N}=
\frac{1}{\det(I_{N}
 - B^*A)^N}.
\end{equation*}
This identity is almost apparent in view of the orthogonality
relation (\ref{or1}).

In the previous section we verified identity (\ref{sint3}), of
which (\ref{7a}) is a special case ($m=n$). Identity (\ref{sint3})
implies
\begin{equation}\label{10aaa}
\int_{U(N)}\frac{d\mu_H(U)}{\det(I_N-AU)^m\det(I_N-U^*B^*)^n}=
\int_{Q^*Q\le I_m}\frac{d\rho_{N,n\times m}(Q)}{\det(I_{mN}
 -Q^*Q \otimes B^*A)}, \hspace{3ex} m,n\le N.
\end{equation}
in the same way as (\ref{7a}) implies (\ref{10aa}). In the range
$m+n\le N$ one can use the explicit expression (\ref{2mleN}) for
$d\rho_{N,n\times m}(Q)$. Replacing the integration over the
$n\times m$ matrices $Q$ by integration over $k\times k$ matrices
$Z$, $k=\min(m,n)$, as in (\ref{2mleN-1}), one obtains the
identity which was claimed in (\ref{jacob}).

Consider now the matrix integral
\begin{equation}\label{10ad}
\int_{U(N)} \det(I_N+AU)^m\det(I_N+U^*B^*)^m\ d\mu_H(U)
=\sum_{\lambda}\
\frac{s_{\lambda}^2(I_m)}{s_{\lambda^{\prime}}(I_N)}\
s_{\lambda^{\prime}} (B^*A)
\end{equation}
where $\lambda^{\prime}$ is the partition conjugate to $\lambda$.
This integral is dual to the one in (\ref{10a}). The equality in
(\ref{10ad}) is a straightforward consequence of the orthogonality
relations (\ref{or1}) and the dual Cauchy identity
(\ref{cauchyid2}) which is a generalization of the well known
expansion of the determinant $\det(I+M)$ in elementary symmetric
functions $e_r$ of the eigenvalues $z_1, \ldots , z_m$ of $M$,
\[
\det(I+M)=\prod_{j=1}^m(1+z_j)=\sum_{r=0}^m e_r(z_1, \ldots , z_m).
\]
In \cite{FK} we proved identity (\ref{14}) which is dual to
(\ref{7a}). It is apparent that (\ref{14}) implies the matrix
integral claimed in (\ref{15}) and vice versa. We shall now give
an independent derivation of (\ref{15}) from the fermionic version
(\ref{fcft}) of the CFT and thus showing that the fCFT implies
(\ref{14}).

Recall that $\det M$ can be written as a Gaussian integral over
anticommuting variables $\varphi_j$, $\varphi_j^*$ with Berezin's
integration rules $ \int (1,\varphi_j) d\varphi_j=(0,1)$, $\int
(1,\varphi^*_j) d\varphi^*_j=(0,1)$,
\[
\det M = \int e^{\sum_{i,j}\varphi_i^* M_{ij}\varphi_j}
\prod_{j}d\varphi_jd\varphi_j^* = \int e^{\
\vec{\varphi^*}M\vec{\varphi}}\ (d\varphi).
\]
By doubling the dimension,
\[
\det(I_N+AU)\det(I_N+U^*B^*)=-\det \left(
                                    \begin{array}{cc}
                                      0 & U^*+A \\
                                      U+B^* & 0 \\
                                    \end{array}
                                  \right)
\]
and
\begin{equation}\label{16}
\det(I_N+AU)^m\det(I_N+U^*B^*)^m = \pm \int \int e^{\tr
\left[\Psi^*(U+B^*)\Phi + \Phi^*(U^*+A)\Psi \right]} (d\Phi)(d\Psi).
\end{equation}
Here $\Phi$ and $\Psi$ are $N\times m$ and $\Phi^*$ and $\Psi^*$
are $m\times N$ matrices with anticommuting entries and $(d\Phi)$
(correspondingly $(d\Psi)$) is the product of the
``differentials'' of the matrix entries of $\Phi$ and $\Phi^*$
(correspondingly, $d\Psi$ and $d\Psi^*$). The sign in front of the
integral in (\ref{16}) depends on the particular ordering of terms
in these products. It is not essential for our calculation (one
can always verify the right sign at the end of calculation) and
will be omitted.

On substituting (\ref{16}) in the integral on the left-hand side
in (\ref{10ad}) and applying the fCFT (\ref{fcft}) one reduces
this integral to the following one
\begin{equation}\label{17}
\int \int \left(\int_{\C^{m\times m}} e^{\tr\left(\Psi^*B^*\Phi
+\Phi^*A\Psi +\Psi^*\Psi Q - Q^*\Phi^*\Phi \right) }
d\mu^{F}_{N,m}(Q) \right) (d\Phi) (d\Psi)
\end{equation}
The quadratic form in the exponential,
\begin{eqnarray*}
\lefteqn{\tr\left(\Psi^*B^*\Phi +\Phi^*A\Psi +\Psi^*\Psi Q -
Q^*\Phi^*\Phi \right) =} \\ & &  \sum_{j=1}^m \left( \vec{\psi^*}_j
B^*\vec{\varphi}_j + \vec{\varphi^*}_j A\vec{\psi}_j \right) +
\sum_{i,j=1}^m \left(Q_{kj}\vec{\psi^*}_j\vec{\psi}_k -
Q^*_{jk}\vec{\varphi^*}_j\vec{\varphi}_k  \right),
\end{eqnarray*}
where $ \vec{\varphi}_j$ and $ \vec{\psi}_j$ are the columns of
$\Phi$  and $\Psi$, and $ \vec{\varphi^*}_j$ and $ \vec{\psi^*}_j$
are the rows of $\Phi^*$ and $\Psi^*$, is the one defined by the
matrix
\[
M= \left(
  \begin{array}{cc}
    -Q^*\otimes I_N & I_m\otimes A \\[2ex]
    I_m\otimes B^* & Q\otimes I_N \\
  \end{array}
\right).
\]
Therefore, $\int \int \ldots (d\Psi)(d\Phi)$ in (\ref{17}) yields
$ \det M = \det (Q^*Q\otimes I_N + I_m\otimes B^*A) $ and one
arrives at the identity
\begin{equation}\label{18}
\int_{U(N)} \hspace{-2ex}\det(I_N+AU)^m\det(I_N+U^*B^*)^m\
d\mu_H(U)=\int_{\C^{m\times m}}\hspace{-2ex} \det (Q^*Q\otimes I_N +
I_m\otimes B^*A) d\mu^{F}_{N,m} (Q).
\end{equation}
On making the substitution
\[
Q=Z^{-1}, \hspace{2ex} (dQ)=\det (Z^*Z)^{2m} (dZ),
\]
in the integral on the right-hand side in (\ref{18}), one obtains
the matrix integral (\ref{15}).

Identity (\ref{7a}) and its dual version (\ref{14}) are rather
useful in the context of Schur function expansions. For example,
the matrix integrals (\ref{19}) and (\ref{20}) are straightforward
corollaries of these identities. Consider, for example, the matrix
integral in (\ref{19}). Expanding each of the determinants on the
left-hand side with the help of (\ref{cauchyid1}), one arrive by
the way of the orthogonality relation (\ref{or4}) at
\[
\int_{Q^*Q\le I_m} \hspace{-0.5ex}
\frac{d\rho_{N,m}(Q)}{\det(I_m-Z_1Q^*)^N\det(I_m-QZ_2^*)^N}=
\sum_{\lambda}\  \frac{s_{\lambda}^2(I_N)}{s_{\lambda}^2(I_m)} \
s_{\lambda} (Z_1Z_2^*)\hspace{-1ex} \int_{Q^*Q\le I_m} \hspace{-2ex}
s_{\lambda} (Q^*Q) d\rho_{N,m\times m}(Q).
\]
In view of (\ref{7a}), the series on the right-hand side folds to
\[
 \sum_{\lambda}\ s_{\lambda}(I_N) \ s_{\lambda}
(Z_1Z_2^*)=\frac{1}{\det(I_m-Z_1Z_2^*)^N}.
\]
Hence (\ref{7a}) implies (\ref{19}) and vice versa. Similarly,
(\ref{14}) implies (\ref{20}) and vice versa.

If $N \ge 2m$ then $d\rho_{N,m\times m}=d\mu^{B}_{N,m}$ and
(\ref{19}) reads
\begin{equation}\label{19a}
\int_{Q^*Q < I_m}
\frac{d\mu^{B}_{N,m}(Q)}{\det(I_m-Z_1Q^*)^N\det(I_m-QZ_2^*)^N}
=\frac{1}{\det(I_m-Z_1Z_2^*)^N}.
\end{equation}

The matrix integrals (\ref{19a}) and (\ref{20}) are variants of
integrals obtained by Berezin in \cite{B1,B2}. We would like to
elaborate more on this link and quote some of Berezin's results.
Consider complex rectangular matrices $Z$ and define
\begin{equation}\label{omega}
\Omega_{B}=\{Z\in \C^{n\times m}: \hspace{1ex} Z^*Z< I_m \}
\hspace{2ex}, \hspace{2ex} \Omega_{F}=\C^{n\times m},
\end{equation}
and, (cf. (\ref{mub}) and (\ref{muf}))
\begin{eqnarray*}
d\mu^{B}_{N,n\times m} (Z)&=& \hbox{const.} \det(I-Z^*Z)^{-n-m+N}\
(dZ),
\hspace{2ex} Z\in \Omega_{B},  \hspace{2ex} N \ge n+m,\\
d\mu^{F}_{N, n\times m} (Z) &=& \hbox{const.} \det(I+Z^*Z)^{-n-m-N}\
(dZ), \hspace{2ex} Z\in \Omega_{F}, \hspace{2ex} N \ge 0,
\end{eqnarray*}
where $(dZ)$ is the cartesian volume element in $\C^{n\times m}$.
The multiplicative constants are fixed by the normalisation
$\int_{\Omega}d\mu_{N, n\times m} (Z)=1$.

Let ${\cal A}_B$ be the Hilbert space of analytic (in the $nm$
variables $Z_{ij}$) functions on $\Omega_B$ with the scalar product
\begin{equation}\label{24}
(f,g) =  \int_{\Omega_{B}} f(Z)\overline{g(Z)}\ d\mu^{B}_{N, n\times
m} (Z)
\end{equation}
and ${\cal A}_F$ be the Hilbert space of analytic (in the $nm$
variables $Z_{ij}$) functions on $\Omega_F$ with the scalar product
\begin{equation}\label{25}
(f,g) =  \int_{\Omega_{F}} f(Z)\overline{g(Z)}\ d\mu^{F}_{N, n\times
m} (Z).
\end{equation}
For $(f,f)$ in (\ref{25}) to be finite, $f(Z)$ cannot grow at
infinity faster than a certain power of $\|Z\|$. Hence ${\cal
A}_F$ consists of polynomials in $Z_{ij}$ and is a
finite-dimensional subspace of $L^2(\Omega_{F}, d\mu^{F}_{N,
n\times m})$. Let us choose a basis $f_j(Z)$ in ${\cal A}_F$.
Given a vector $f$ in $L^2(\Omega_{F}, d\mu^{F}_{N, n\times m})$,
its orthogonal projection on ${\cal A}_F$ is given by
\[
(K_Ff)(Z)=\sum_{j} (f,f_j)f_j(Z) =\int_{\Omega_F} K_F(Z,\bar Q) f(Q)
d\mu^{F}_{N, n\times m}(Q),
\]
where $K_F(Z,\bar Q)=\sum_{j}f_j(Z)f_j(\bar Q)$. As $K_F(Z,\bar Q)$
is the kernel of the operator of orthogonal projection onto ${\cal
A}_F$, it is independent of the choice of basis there. Similarly, if
$f_j$ is a basis in ${\cal A}_B$ then $K_B(Z,\bar
Q)=\sum_{j}f_j(Z)f_j(\bar Q)$ is the kernel of the orthogonal
projection from $L^2(\Omega_{B}, d\mu^{B}_{N, n\times m})$ onto
${\cal A}_B$. The Hilbert space ${\cal A}_B$ is infinite dimensional
and, hence, there is a question of convergence of the series for
$K_B(Z,\bar Q)$. It can be shown that this series converges
absolutely and uniformly on compact sets, see \cite{B1}. The kernels
$K_i(Z,\bar Q)$ define coherent states in ${\cal A}_i$, $i=B,F$.
Indeed, for each $Q\in \Omega_i$, $K_i(Z, \bar Q)$ as a function in
$Z$, $f_{\bar Q}(Z)=K_i(Z, \bar Q)$, belongs to ${\cal A}_i$ and for
any $f$
\begin{equation}\label{29}
(f,f_{\bar Q})=\sum_{j} (f,f_j)f_j(Q)=f(Q).
\end{equation}
Hence, one has the resolution of identity
\[
(f,g)= \int_{\Omega_i} (f,f_{\bar Q}) (f_{\bar Q}, g) d\mu^{i}_{N,
n\times m}(Q), \hspace{3ex} i=B,F.
\]
The vectors $f_{\bar Q}$ are not orthogonal. Putting $f_{\bar Z}$
for $f$ in (\ref{29}), one obtains $(f_{\bar Z}, f_{\bar Q})=f_{\bar
Z}(Q)$, or, equivalently
\begin{equation}\label{26}
\int_{\Omega_i} K_i(Q,\bar R) K_i(R,\bar Z)d\mu^{i}_{N, n\times
m}(R)=K_i(Q,\bar Z) \hspace{3ex} i=B,F,
\end{equation}
which means that the kernels $K_i(Z,\bar Q)$ are reproducing. The
coherent states which are defined by these kernels played an
important role in Berezin's construction of quantization in
symmetric spaces. For some parameter values these kernels can be
found in the explicit form \cite{B2}
\begin{eqnarray*}
K_B(Z,\bar Q) &=& \det(I_m-Q^*Z)^{-N} \hspace{2ex} \hbox{for real
$N\ge n+m$,} \\ K_F(Z,\bar Q)&=& \det(I_m-Q^*Z)^{N} \hspace{3.3ex}
\hbox{for $N=0,1,2, \ldots $,}
\end{eqnarray*}
and identity (\ref{26}) expressing the reproducing property of these
kernels takes the form of the matrix integrals
\begin{equation}\label{27}
\int_{\Omega_B} \frac{d\mu^{B}_{N, n\times
m}(Q)}{\det(I_m-Q^*Z_1)^N\det(I_m-Z_2^*Q)^N}
=\frac{1}{\det(I_m-Z_2^*Z_1)^N}
\end{equation}
and
\begin{equation}\label{28}
\int_{\Omega_F} \det(I_m+Q^*Z_1)^N\det(I_m+Z_2^*Q)^N\ d\mu^{F}_{N,
n\times m}(Q) = \det(I_m+Z_2^*Z_1)^N,
\end{equation}
of which (\ref{19}) and (\ref{20}) are particular cases.
Correspondingly, the matrix integrals (\ref{27}) and (\ref{28})
allows one to extend identities (\ref{7a}) and (\ref{14}) to
rectangular matrices as claimed in (\ref{sint3}) and (\ref{14a}).
Indeed, assuming that $N$ is a non-negative integer and
expanding the integrands in (\ref{27}) and (\ref{28}) in Schur
functions by the means of the Cauchy identities
(\ref{cauchyid1})--(\ref{cauchyid2}), and then applying the
orthogonality relation (cf. (\ref{or4}))
\begin{equation}\label{add6}
\int_{\Omega_i} s_{\lambda} (AQ) s_{\mu} (Q^*B^*)\ d\mu^{i}_{N,
n\times m} (Q)=\delta_{\lambda,\mu}\frac{s_{\lambda}
(AB^*)}{s_{\lambda}(I_n)s_{\lambda}(I_m)} \int_{\Omega_i}
s_{\lambda} (Q^*Q)\ d\mu^{i}_{N, n\times m} (Q), \hspace{3ex}
i=B,F,
\end{equation}
one obtains Schur function expansions for the determinantal powers
on the right-hand side in (\ref{27}) and (\ref{28}). On comparing
the coefficients in these expansions with those in
(\ref{cauchyid1})--(\ref{cauchyid2}) one arrives at (\ref{sint3})
and (\ref{14a}).

Identities (\ref{sint3}) and (\ref{14a}) can also be derived in an
elementary way and independently of (\ref{27}) and (\ref{28}) and
then used to prove (\ref{27}) and (\ref{28}). We shall demonstrate
this at the end of this section.

With identity (\ref{14a}) in hand, we can revisit the matrix
integral (\ref{10ad}) and consider unequal powers of determinants
\begin{equation}\label{10ad-1}
\int_{U(N)} \det(I_N+AU)^m\det(I_N+U^*B^*)^n\ d\mu_H(U)
=\sum_{\lambda}\ \frac{s_{\lambda}
(I_m)s_{\lambda}(I_n)}{s_{\lambda^{\prime}}(I_N)}\
s_{\lambda^{\prime}} (B^*A).
\end{equation}
By making use of (\ref{14a}), one can fold the sum on the right,
\[
\sum_{\lambda}\ \frac{s_{\lambda}
(I_m)s_{\lambda}(I_n)}{s_{\lambda^{\prime}}(I_N)}\
s_{\lambda^{\prime}} (B^*A)= \int_{\C^{n\times m}}
\det(I+Z^*Z\otimes B^*A)\ d\mu^{F}_{N,n\times m}(Z).
\]
Replacing the integration over the $n\times m$ matrices $Z$ by
integration over $k\times k$ matrices $Z$, $k=\min(m,n)$, as in
(\ref{2mleN-1}), one obtains the identity claimed in
(\ref{jacobdual}).

Now we turn to identities (\ref{sint3}) and (\ref{14a}). The
integration on the left-hand side in (\ref{sint3}) and (\ref{14a})
goes effectively over the eigenvalues of $Q^*Q$. Consider $N\ge
2m$ assuming without loss of generality that $m\le n$. Recalling
the singular value decomposition for $Q$, $Q=H\sqrt{X}V^*$, where
$V$ is $m\times m$ unitary, i.e. $V\in U(m)$, $H$ is $n\times m$
unitary, i.e. $H\in V_m(\C^n)$ (see Section 2), and $X$ is
diagonal $m\times m$ matrix of the eigenvalues $x_1, \ldots, x_m$
of $Q^*Q$, one can make the substitution $Q=H\sqrt{X}V^*$ in the
integrals in (\ref{sint3}) and (\ref{14a}). The corresponding
Jacobian is well known, see e.g. \cite{Muirhead,Mathai},
\[
(dQ)=\hbox{const.}\prod_{1\le i<k \le m} (x_i-x_j)^2 \prod_{i=1}^m
x^{(n-m)}_j \prod_{j=1}^m dx_j (H^*dH)(V^*dV),
\]
where $(H^*dH)$ and $(V^*dV)$ are the invariant volume elements
in, respectively $V_m(\C^n)$ and $U(m)$. This substitution reduces
the matrix integral in  (\ref{sint3}) to (\ref{30}) and the one in
(\ref{14a}) to (\ref{31}). The normalization constants are given
by
\begin{eqnarray}\label{32}
c^N_{n,m} &=& \int_0^1 \hspace{-1ex}\cdots \int_0^1 \prod_{j=1}^m\
x_j^{n-m} (1-x_j)^{N-m-n}\hspace{-1ex} \prod_{1\le i<j\le
m}\hspace{-1.5ex}(x_i-x_j)^2 \ \prod_{j=1}^m dx_j  \\ \nonumber
&=&
\prod_{j=0}^{m-1}\frac{(1+j)!\,(n-m+j)!\,\Gamma(N-n-m+j+1)}{\Gamma(N-m+j+1)}
\end{eqnarray}
and
\begin{eqnarray}\label{33}
k^N_{n,m}&=&\int_0^{\infty} \hspace{-1.5ex} \cdots \int_0^{\infty}
\prod_{j=1}^m\ \frac{x_j^{n-m}}{ (1+x_j)^{N+m+n}} \prod_{1\le
i<j\le m}\hspace{-1.5ex}(x_i-x_j)^2\ \prod_{j=1}^mdx_j\\ \nonumber
&=& \prod_{j=0}^{m-1}\frac{(1+j)!\,(n-m+j)!\,\Gamma(N+j+1)}{\Gamma
(N+n+j+1)}.
\end{eqnarray}
The integrals in (\ref{32}) and (\ref{33}) are particular cases of
the Selberg integral
\begin{eqnarray*}
\int_0^1 \hspace{-1ex}\cdots \int_0^1 \prod_{j=1}^m  x_j^{p-1}
(1-x_j)^{q-1}\hspace{-2ex} \prod_{1\le i<j\le
m}\hspace{-1.5ex}|x_i-x_j|^{2\gamma} \prod_{j=1}^mdx_j =
\prod_{j=0}^{m-1}\frac{\Gamma (p+j\gamma)\Gamma (q+j\gamma) \Gamma
(1+(1+j)\gamma)}{\Gamma (p+q+(m+j-1)\gamma)\Gamma (1+\gamma)} \\ =
 \int_0^{\infty} \hspace{-1.5ex} \cdots
\int_0^{\infty}\prod_{j=1}^m  \frac{t_j^{p-1}}{
(1+t_j)^{p+q+2(m-1)\gamma}} \prod_{1\le i<j\le m}\hspace{-1.5ex}
|t_i-t_j|^{2\gamma} \prod_{j=1}^mdt_j, & &
\end{eqnarray*}
which is a multivariate generalization of the Euler beta integral
\[
\int_{0}^1 x^{p-1}(1-x)^{q-1}\ dx=\int_0^{\infty}\frac{t^{p-1}\
dx}{(1+t)^{p+q}}=\frac{\Gamma (p) \Gamma (q)}{\Gamma (p+q)}=B(p,q).
\]
Here $\Gamma$ and $B$ are the Gamma and Beta functions,
respectively.

The rest of this section is devoted to explicit evaluation of the
integral in (\ref{30})  and the one in (\ref{31}). To the best of
our knowledge, this evaluation is new.

Let
\[
S^B_{\lambda}(p,q;m)=\int_0^1 \hspace{-1ex}\cdots \int_0^1
s_{\lambda}(x_1, \ldots, x_m)  \prod_{j=1}^m\ x_j^{p-1}
(1-x_j)^{q-1}\hspace{-1ex} \prod_{1\le i<j\le
m}\hspace{-1.5ex}(x_i-x_j)^2 \ \prod_{j=1}^m dx_j
\]
and
\[
S^F_{\lambda}(p,q;m)=\int_0^{\infty} \hspace{-1.5ex} \cdots
\int_0^{\infty} s_{\lambda}(x_1, \ldots, x_m) \prod_{j=1}^m\
\frac{x_j^{p-1}}{(1+x_j)^{p+q+2(m-1)}} \hspace{-1ex} \prod_{1\le
i<j\le m}\hspace{-1.5ex}(x_i-x_j)^{2} \ \prod_{j=1}^m dx_j
\]
In view of (\ref{schur}),
\[
s_{\lambda}(x_1, \ldots, x_m) \prod_{1\le i<j\le
m}\hspace{-1.5ex}(x_i-x_j)^2 =
{\det\left(x_i^{m+\lambda_j-j}\right)_{i,j=1}^m}{\det\left(x_i^{m-j}\right)_{i,j=1}^m}.
\]
By making use of the Gram identity
\begin{equation}\label{Gram}
\int \cdots \int \det \left(F_i(x_j)\right)_{i,j=1}^m \det
\left(G_i(x_j)\right)_{i,j=1}^m \prod_{i=1}^m dx_i = m! \det \left(
\int F_i(x) G_j(x) dx \right)_{i,j=1}^m,
\end{equation}
we have
\[
S^B_{\lambda}(p,q;m) =m!\det\left(\int_0^1
x^{m+p+f_j-i-1}(1-x)^{q-1}dx\right)_{i,j=1}^m\hspace{-2ex}= m! \det
\left(B(m+p+f_j-i,q)\right)_{i,j=1}^m,
\]
where we have introduced $f_j=m+\lambda_j-j$, and
\begin{eqnarray}\nonumber
S^F_{\lambda}(p,q;m)&=& m!\det\left(\int\limits_0^{\infty}
\frac{x^{m+p+f_j-i-1}dx}{(1-x)^{p+q+2(m-1)}}\right)_{i,j=1}^m\hspace{-2ex}\\
\label{36} &=& m!\det\left( B(m+p+f_j-i,
q+m-f_j-2+i)\right)_{i,j=1}^m
\end{eqnarray}
In \cite{FK} we proved the following identity for binomial
determinants
\begin{equation}\label{37}
\det\left(B(p_j-i,q_j+i)\right)_{i,j=1}^m=\det\left(B(p_j-i,q_j+1)\right)_{i,j=1}^m.
\end{equation}
By making use of this identity,
\begin{eqnarray*}
S^B_{\lambda}(p,q;m) &=& m! \det \left(B(m+p+f_j-i,q+i-1)\right)_{i,j=1}^m\\
&=& m!\left\{\prod_{j=1}^{m} \frac{\Gamma (q+i-1)}{\Gamma
(m+p+q+f_j-1)}\right\}\det \left (\Gamma
(m+p+f_j-i)\right)_{i,j=1}^m
\end{eqnarray*}
Now, recalling the standard determinant
\begin{equation}\label{38}
\det\left( \Gamma (p_j+m-i) \right)_{i,j=1}^m=\prod_{j=1}^m \Gamma
(p_j)\hspace{-1ex} \prod_{1\le i<j\le m} (p_i-p_j),
\end{equation}
one arrives at
\begin{equation}\label{add7}
S^B_{\lambda}(p,q;m)= m!\left\{\prod_{j=1}^{m} \frac{\Gamma
(q+j-1)\Gamma (p+f_j)}{\Gamma (m+p+q+f_j-1)}\right\} \hspace{-1ex}
\prod_{1\le i<j\le m} (f_i-f_j), \qquad f_j=m+\lambda_j-j.
\end{equation}
Similarly, applying (\ref{36}) to the determinant on the right in
(\ref{36}), one obtains
\begin{eqnarray*}
S^F_{\lambda}(p,q;m)&=&m! \det \left(B(m+p+f_j-i,q+m-f_j-1)\right)_{i,j=1}^m\\
&=& m!\left\{\prod_{j=1}^{m} \frac{\Gamma (q+m-f_j-1)}{\Gamma
(p+q+2m-j-1)}\right\}\det \left(\Gamma (m+p+f_j-i)\right)_{i,j=1}^m,
\end{eqnarray*}
and, by (\ref{38}),
\begin{equation}\label{add8}
S^F_{\lambda}(p,q;m)= m!\left\{\prod_{j=1}^{m} \frac{\Gamma
(p+f_j)\Gamma (q+m-f_j-1) }{\Gamma
(p+q+2m-j-1)}\right\}\hspace{-1ex} \prod_{1\le i<j\le m}
(f_i-f_j), \qquad f_j=m+\lambda_j-j.
\end{equation}

We would like to finish this section with a calculation showing
that (\ref{add7}) implies (\ref{27}). As has been mentioned above,
for integer $N$ this claim follows from the Cauchy identity
(\ref{cauchyid1}). The case of non-integer $N$ can be handled with
the help of the following identity where $M$ is $m\times m$:
\begin{equation}\label{add9}
\frac{1}{\det (I_m -M)^N}=\sum_{\lambda} \beta^B_{\lambda}
s_{\lambda}(I_m) s_{\lambda}(M), \quad
\beta^B_{\lambda}=\prod_{j=1}^m \frac{\Gamma (N+\lambda_j-j+1)
(m-j)!}{\Gamma (N-j+1) (m+\lambda_j-j)!},
\end{equation}
This identity is due to Hua \cite{Hua}. In Appendix we derive it
and its dual version with the help of integration over the unitary
group, see examples 4 and 5.

It can be verified from the Selberg integral (\ref{add7}) that for
real $N\ge n+m$ and $n\ge m$
\begin{equation}\label{add11}
\int_{Q^*Q\le I_m} s_{\lambda} (Q^*Q)\ d\mu^B_{N,n\times m} (Q) =
\frac{s_{\lambda} (I_n)}{\beta^B_{\lambda}}.
\end{equation}
By making use of the Schur function expansion in (\ref{add9}) and
orthogonality relation (\ref{add6}), the left-hand side in
(\ref{27}) can be expanded as follows
\[
\sum_{\lambda} \frac{(\beta^B_{\lambda})^2
s_{\lambda}(I_m)}{s_{\lambda}(I_n)} \left(\int_{Q^*Q\le I_m}
s_{\lambda} (Q^*Q)\ d\mu^{B}_{N, n\times m} (Q)\right) s_{\lambda}
(Z_2^*Z_1).
\]
Applying now (\ref{add11}) and then folding the series with the
help of (\ref{add9}), one gets the right-hand side of (\ref{27}).
Hence (\ref{add9}) implies (\ref{27}) for real $N\ge n+m$.

A similar calculation shows that (\ref{add8}) implies (\ref{28})
for integer $N\ge 0$. One only needs to recall the dual Cauchy
identity (\ref{cauchyid2}).

\section{Deformed version of the bosonic CFT.}
\label{section5} In order to introduce the deformation of the
bosonic CFT that we are going to derive, we would like to
demonstrate an evaluation of the right-hand side of the bosonic
CFT in terms of the eigenvalues of the matrix $X^*XY^*Y$. We only
consider the range $2m\le N$, however, a similar calculation gives
an answer in the range $N<2m<2N$. As we shall eventually replace
the integration over the matrix ball which is a bounded domain by
integration over a hyperbolic domain which is unbounded, we first
rewrite (\ref{bcft-xy1}) by changing to complex exponentials
\begin{equation}\label{bcft4}
\int_{U(N)}e^{-i\tr (Y^*UX + X^*U^{*}Y) }d\mu_H(Q) =\hbox{const.}
\int_{Q^{*}Q\le I_m} e^{-i\tr (X^*X Q + Q^* Y^*Y ) } \det
(I-Q^*Q)^{N-2m} (dQ).
\end{equation}

Denote $Q_A=X^*X$ and $Q_R=Y^*Y$ and let $D$ is diagonal matrix of
the square roots of the eigenvalues of the matrix product $Q_AQ_R$.
Then one can always find a non-degenerate matrix $T$ such that
\begin{equation}\label{tdt}
Q_A=TDT^* \qquad \hbox{and} \qquad Q_R=(T^*)^{-1} D T^{-1}.
\end{equation}
Such parametrisation is possible for any two positive definite
matrices \cite{FS}. This can be seen by diagonalising the matrix
$Q_R^{1/2}Q_AQ_R^{1/2}$. Writing $Q_R^{1/2}Q_AQ_R^{1/2}=UDU^*$, we
have $T=Q_R^{-1/2}UD^{1/2}$. It is apparent from (\ref{tdt}) that
$T$ is defined up to the right multiplication by diagonal unitary
matrices.

By making use of (\ref{tdt}) and singular value decomposition for
$Q$
\[
Q=U\diag (q_1, \ldots, q_m) V,\qquad (dQ)\propto (U^*dU)(V^*dV)
\prod_{1\le j<k\le m} (q_j^2-q_k^2)^2\prod_{j=1}^m d(q_j^2),
\]
one rewrites the integral on the right-hand side in (\ref{bcft4}) as
follows
\begin{eqnarray}\nonumber
\lefteqn{\int_0^1 \ldots \int_0^1 \prod_{j=1}^m \ dq_j\, q_j\,
(1-q_j^2)^{n-2m}\hspace{-2ex} \prod_{1\le j<k\le m}
(q_j^2-q_k^2)^2\times } \\ \label{isw} & &
\int_{U(m)}d\mu_H(U)\int_{U(m)}d\mu_H(V)\
 \exp\left[-i\tr (U\hat q V TDT^* +
(T^*)^{-1}DT^{-1}V^*\hat q U^*\right],
\end{eqnarray}
where $\hat q=\diag (q_1, \ldots q_m)$. The integral in (\ref{isw})
was computed in \cite{SW},
\[
\int_{U(m)}\hspace{-2ex}d\mu_H(U)\int_{U(m)}\hspace{-2ex}d\mu_H(V)\
 e^{\frac{1}{2}\tr (UAVB +
CV^*DU^*)} \propto \frac{\det \big(I_0(x_jy_k)\big)_{j,k=1}^m
}{\prod_{1\le j<k\le m} (x_j^2-x_k^2) \prod_{1\le j<k\le m}
(y_j^2-y_k^2)}
\]
where $x_j^2$ and $y_j^2$ are the eigenvalues of $AB$ and $BC$,
respectively, and $I_0$ is the modified Bessel function of zero
order. In our case $AD=-4\hat q^2$ and $BC=TD^2T^{-1}$, and
collecting everything together we obtain
\begin{eqnarray*}
\lefteqn{\int_{U(N)}e^{-i\tr (Y^*UX + X^*U^{*}Y) }d\mu_H(Q)=} \\
\nonumber & &\frac{\hbox{const.}}{\prod\limits_{1\le j<k\le
m}(d_j^2-d_k^2)} \int_0^1 \ldots \int_0^1 \prod_{j=1}^m \ dq_j q_j
(1-q_j^2)^{n-2m}\hspace{-2ex} \prod_{1\le j<k\le m} (q_j^2-q_k^2)
\det \big(J_0(2q_jd_k)\big)_{j,k=1}^m,
\end{eqnarray*}
where $J_0(z)$ is the Bessel function of zero order. By making use
of the Gram formula (\ref{Gram}),
\begin{equation}\label{45}
\int_{U(N)}e^{-i\tr (Y^*UX + X^*U^{*}Y) }d\mu_H(Q)\ \propto \
\frac{\det \left( \int\limits_0^1  J_0(2qd_k)
q^{2(m-j)+1}(1-q^2)^{N-2m}dq \right)_{j,k=1}^m }{\prod\limits_{1\le
j<k\le m}(d_j^2-d_k^2)}.
\end{equation}
This formula can also be derived directly from the character
expansion (\ref{40-2}), see Lemma 5 in \cite{FK}. Such calculation
is standard, and in fact the above two-fold integral over unitary
group was evaluated in \cite{SW} using the character expansion
method \cite{Balantekin}. Formula (\ref{45}) can also be extended to
the range $N<2m<2N$. In this case one gets the Bessel function $J_0$
and its derivatives in the determinant on the right-hand side in
(\ref{45}).

Formula (\ref{45}) allows one to obtain an alternative version of
the bosonic CFT  with the integration manifold in the right-hand
side parametrized by the matrices $Q_1$ and $Q_2$ defined in
Eq.(\ref{q1q2}). Now, consider the integral
\begin{eqnarray}\nonumber
F_m(Q_A,Q_R)&=&\int (dQ_1dQ_2) \det (I_m-Q_1Q_2)^{N-2m} \exp
\left[-i\tr (Q_1Q_R+Q_2Q_A)\right]\\ \nonumber &=& \hbox{const.}
\int_{-1}^1\ldots \int_{-1}^1 \prod_{1\le j<k\le m}(p_j^2-p_k^2)^2
\prod_{j=1}^m p_j(1-p_j^2)^{N-2m}\prod_{j=1}^m dp_j \\ \label{46} &
& \times \int_{GL_m(\C)}d\mu_H(T) \exp \left\{-i\left[TPT^*Q_R
+(T^*)^{-1}PT^{-1}Q_A \right] \right\}
\end{eqnarray}
Recalling the $Q_A=T_0DT_0^*$ and $Q_R=(T_0^*)^{-1}DT_0^{-1}$ for
some $T_0\in GL_m(\C)$, see (\ref{tdt}), we can rewrite the integral
in (\ref{46}) in terms of $P$, $D$ and $T$. The matrix $T_0$
disappears because of the invariance of $d\mu_{H}(T)$. The resulting
integral is known \cite{FS,F2}
\[
\int_{GL_m(\C)}\hspace{-2ex}d\mu_H(T) \exp \left\{-i\left[T^*DT
+T^{-1}D(T^*)^{-1} \right]P \right\} = \frac{\hbox{const.}\det
\big(K_0(2ip_jd_k)\big)_{j,k=1}^m}{\prod_{1\le j<k\le m}
(p_j^2-p_k^2)\prod_{1\le j<k\le m} (d_j^2-d_k^2) }
\]
where $K_0(z)$ is the Macdonald function. Note that because the
matrices $D$ and $P$ are diagonal, the above integral is effectively
going over the right cosets $GL_m(\C)/U(1)\times \ldots \times
U(1)$. On substituting this expression back in (\ref{46}) and using
the Gram formula (\ref{Gram}), one obtains
\begin{equation}\label{47}
F_m(Q_A,Q_R)\ \propto \ \frac{\det \left( \int\limits_{-1}^1
K_0(2pd_k) p^{2(m-j)+1}(1-p^2)^{N-2m}dp \right)_{j,k=1}^m
}{\prod\limits_{1\le j<k\le m}(d_j^2-d_k^2)}.
\end{equation}
Recall the identity (\cite{GR}, Equations 8.405 and 8.421)
\[
K_0(iu)=-\frac{\pi}{2}\left[Y_0(|u|)+i\sgn (u) J_0(|u|)  \right]
\]
where $Y_0$ is the Neumann function. It follows from this identity
that
\[
\int_{-1}^1 K_0(2pd) p^{2(m-j)+1}(1-p^2)^{N-2m}dp = 2i \int_0^1
J_0(2pd) p^{2(m-j)+1}(1-p^2)^{N-2m}dp.
\]
Therefore, the determinants in (\ref{45}) and (\ref{47}) differ
only by a multiplicative constant, and we arrive at the variant of
the bosonic CFT , Eq.(\ref{cft5}).

It is instructive to write the formula (\ref{cft5}) for the
simplest but yet non-trivial case of $m=1$. In this case $Q_1$ and
$Q_2$ are just real numbers,
\[
Q_1=|t|^2p, \quad Q_2=\frac{p}{|t|^2}, \quad p\in [-1,1], \quad
t=Re^{i\theta}\in \C,
\]
the integration measure is
\[
(dQ_1dQ_2)=p(1-p^2)^{N-2}dp\ \frac{dtd\bar t}{|t|^2}=
p(1-p^2)^{N-2}\ dp \ \frac{d R }{R}\ d\theta,
\]
and (\ref{cft5}) reads
\[
\int_{U(N)}\hspace{-2ex}\exp\left[-i\tr (\vec{y}^*U\vec{x} +
\vec{x}^*U^{*}\vec{y})\right] d\mu_H(U)\ \propto\ \int_{0}^1
p(1-p^2)^{N-2}dp \int_{-\infty}^{\infty} \frac{dR}{R}\ \exp
\left[-ip \left(R|\vec{x}|^2+\frac{|\vec{y}|^2}{R}\right)\right].
\]
This should be compared to the original version of the bosonic CFT.
For $m=1$ it reads
\begin{eqnarray*}
\int_{U(N)}\hspace{-2ex}\exp\left[-i\tr (\vec{y}^*U\vec{x} +
\vec{x}^*U^{*}\vec{y})\right] d\mu_H(U) & \propto & \int_{|z|^2\le
1} dz d\bar z (1-|z|^2)^{N-2} \exp \left[-i (z|\vec{x}|^2+\bar
z|\vec{y}|^2)\right] \\ &\propto & \int_{0}^1 p(1-p^2)^{N-2}dp\
J_0(2p|\vec{x}||\vec{y}|).
\end{eqnarray*}
Therefore the transition from the original version of the bosonic
CFT to its deformed version amounts to replacing the Bessel function
by an integral,
\[
J_0(2pab)=\frac{1}{2\pi i} \int_{-\infty}^{\infty}  \frac{dR}{R}\
\exp \left[-ip \left(Ra^2+\frac{b^2}{R}\right)\right], \qquad p,a,b
>0.
\]

We would like to demonstrate the usefulness of the deformed CFT
(\ref{47}) on the example of the matrix integral
\begin{equation}\label{48}
R_{\varepsilon} (BB^*)= \int_{U(N)} \frac{d\mu_H(U)}{\det
[\varepsilon^2 I_N +(I_N-BU)(I_N-U^*B^*)]^m}.
\end{equation}
For $\varepsilon >0$ this integral is well defined for any matrix
$B$. By doubling matrix dimension,
\[
\det[\varepsilon^2 I + (I-UB)(I-U^*B^*)]=\left|\begin{array}{cc}
    \varepsilon I & i(U^*-B) \\
  i(U -B^*) & \varepsilon I
\end{array} \right|
\]
The quadratic form $\vec{f}^* M \vec{f}$,
$\vec{f}^*=(\vec{x}^*,\vec{y}^*)$, corresponding to the $2N\times
2N$ matrix on the right-hand side is
\[
\varepsilon (\vec{x}^*\vec{x} +
\vec{y}^*\vec{y})+i\vec{x}^*B\vec{y}+i\vec{y}^*B^*\vec{x}
-i\vec{x}^*U^*\vec{y}-i\vec{y}^*U\vec{x}.
\]
By making use of the Gaussian integral
\[
\frac{1}{\det M} = \frac{1}{\pi^{2N}}\int_{\C^{2N}} e^{-\vec{f}^* M
\vec{f}}\  (d\vec{f}),
\]
we have
\[
R_{\varepsilon} (BB^*)\ \propto \ \int_{\C^{N\times
m}}\hspace{-1ex}(dX) \int_{\C^{N\times m}}\hspace{-1ex}(dY)
e^{-\varepsilon \tr (X^*X+ Y^*Y -iX^*BY-iY^*B^*X)} J_{cf}
(X^*X,Y^*Y)
\]
where
\[
J_{cf}(X^*X,Y^*Y) = \int_{U(N)} d\mu_H(U) e^{-i\tr (Y^*UX+X^*U^*Y)}
\]
This integral is exactly the one appearing on the left-hand side in
the bosonic CFT (\ref{bcft}). However, if one mindlessly applies
(\ref{bcft}), one gets a diverging integral and the use of CFT for
evaluation of integral (\ref{48}) appears to be problematic. The
deformed version (\ref{cft5}), as we shall show below, is free of
this problem.

By making use of (\ref{cft5}),
\begin{equation}\label{49}
R_{\varepsilon} (BB^*)\ \propto \ \int_{\C^{N\times
m}}\hspace{-1ex}(dX) \int_{\C^{N\times m}}\hspace{-1ex}(dY)
\exp\left[- (\vec{x}^*,\vec{y}^*){\cal M} \left(
                                \begin{array}{c}
                                  \vec{x} \\
                                  \vec{y} \\
                                \end{array}
                              \right),
 \right]
\end{equation}
with the quadratic form in the exponential being
\[
\varepsilon \sum_{j=1}^m (\vec{x}_j^*\vec{x}_j+
\vec{y}_j^*\vec{y}_j)
 -i\sum_{j=1}^m (\vec{x}_j^*B\vec{y}_j+ \vec{y}_j^*B^*\vec{x}_j) + i
 i \sum_{j,k=1}^m (Q_1)_{jk}\vec{x}_k^*\vec{x}_j +i
 \sum_{j,k=1}^m (Q_2)_{jk}\vec{y}_k^*\vec{y}_j.
\]
The matrix $\cal M$ corresponding to this form is
\[
{\cal M}=\left[\begin{array}{cc}
           \varepsilon (I_m +iQ_1)\otimes I_N & -iI_m\otimes B
           \\[1ex]
           -iI_m\otimes B^* & \varepsilon (I_m +iQ_2)\otimes I_N
         \end{array}\right] = \varepsilon I + i \left[\begin{array}{cc}
           Q_1 \otimes I_N & -I_m\otimes B
           \\[1ex]
           -I_m\otimes B^* &  Q_2 \otimes I_N
         \end{array}\right].
\]
Since ${\cal M} =\varepsilon +i\, \hbox{(Hermitian matrix)}$, the
Gaussian integral in (\ref{49}) converges for $\varepsilon >0$ and
is equal to
\[
\frac{1}{\det {\cal M}}= \frac{1}{\det \left\{\left[ (\varepsilon
I_m +iQ_1)(\varepsilon I_m +iQ_2)\right]\otimes I_N + I_m \otimes
B^*B \right\}}
\]
or, on substituting $Q_1=TPT^*$, $Q_2=(T^*)^{-1}PT^{-1}$,
\[
\frac{1}{\det {\cal M}}= \frac{1}{\det \left\{ [\varepsilon
(TT^*)^{-1} +iP][\varepsilon (TT^*) +iP]\otimes I_N + I_m \otimes
B^*B \right\}}.
\]
Thus finally
\begin{eqnarray*}
\lefteqn{\int_{U(N)} \frac{d\mu_H(U)}{\det [\varepsilon^2 I_N
+(I_N-BU)(I_N-U^*B^*)]^m} =  \int_{[-1,1]^m} \prod_{j=1}^m dp_j \
p_j (1-p_j^2)^{N-2m} \times }\\ & &\prod_{1\le j<k\le m}
(p_j^2-p_k^2)^2 \int_{GL_m(\C)} \frac{(dT)}{\det
(T^*T)^m}\prod_{j=1}^N \frac{1}{\det \left\{ [\varepsilon
(TT^*)^{-1} +iP][\varepsilon (TT^*) +iP] + b_j^2 \right\}},
\end{eqnarray*}
where $P=\diag (p_1, \ldots , p_m)$ and $b_j^2$ are the eigenvalues
of $B^*B$.

\bigskip

{\bf Acknowledgements}. This research was in part accomplished
during the first author's stay as a Bessel awardee at the Institute
of Theoretical Physics, University of Cologne, Germany. Y.V.F. is
grateful to M. Zirnbauer for the kind hospitality extended to him
during his months in Cologne and for his interest in this work. The
Humboldt Foundation is acknowledged for the financial support of
that visit. The research in Nottingham was supported by EPSRC grant
EP/C515056/1 "Random Matrices and Polynomials: a tool to understand
complexity". The authors would like to thank Y Wei for useful
communications, and in particular for drawing their attention to
formula (\ref{yw}).

\appendix

\section{Appendix}

In this appendix we would like to demonstrate a simple
calculation, reminiscent of the one given in Section 3 in
\cite{HO}, to determine coefficients in the Schur function
expansions for the class of symmetric functions $g(z_1, \ldots,
z_m)=\prod_{j=1}^m h(z_m)$, where $h(z)$ is analytic in a
neighbourhood of $|z|=1$,
\begin{equation}\label{a1}
g(z_1, \ldots, z_m)=\sum_{\lambda} c_{\lambda} s_{\lambda} (z_1,
\ldots, z_m).
\end{equation}
Thinking of the $z_j$'s as of eigenvalues of unitary matrix $U$, we
can rewrite (\ref{a1}) as $g(U)= \sum_{\lambda} c_{\lambda}
s_{\lambda} (U)$. Then, because of the orthogonality of the Schur
functions on $U(m)$, see (\ref{or1}), the coefficients $c_{\lambda}$
are just the ``Fourier''-coefficients of the function $g(U)$:
\begin{equation}\label{a2}
c_{\lambda}=\int_{U(m)} g (U) \overline{s_{\mu} (U)}\ d\mu_H(U).
\end{equation}
The unitary matrix $U$ can be brought to diagonal form by a unitary
transformation, $U=Ve^{i\Phi}V^*$, where $\Phi = \diag (\phi_1,
\ldots, \phi_m), 0\le \phi < 2\pi$. Correspondingly, the volume
element in $U(m)$ transforms as follows (see e.g. \cite{Hua} or
\cite{Forbook})
\[
(U^*dU)=\hbox{const.}\, \prod_{1\le j<k\le m}
|e^{i\phi_j}-e^{i\phi_k}|^2 \prod_{j=1}^m d\phi_j\ (V^*dV),
\]
and the integral on the right-hand side in (\ref{a2}) reduces to
\[
c_{\lambda}=\frac{1}{m!(2\pi)^m}\int_{0}^{2\pi} \ldots
\int_{0}^{2\pi} \prod_{j=1}^m h(e^{i\phi_j})\ \det
\left(e^{-i\phi_j(m+\lambda_k-k)}\right) \det \left(e^{i\phi_j(m-k)}
\right)\ \prod_{j=1}^m d\phi_j,
\]
where we have also used (\ref{schur}). By the Gram identity
(\ref{Gram}),
\[
c_{\lambda}=\det \left(\frac{1}{2\pi}\int_{0}^{2\pi}
h(e^{i\phi})e^{-i\phi (\lambda_k-k+j)}\ d\phi \right)_{j,k=1}^m.
\]
In other words,
\begin{equation}\label{a5}
c_{\lambda} = \det (\alpha_{\lambda_k-k+j})_{j,k=1}^m, \qquad
\alpha_r = \frac{1}{2\pi}\int_{0}^{2\pi} h(e^{i\phi})e^{-i\phi r}\
d\phi.
\end{equation}

We  would like to give several examples.

\medskip

\emph{Example 1.} Consider the function $g(z_1, \ldots, z_m)=\exp
(\sum_j z_j)$. Then $h(z)=e^{z}$. Expanding the exponential in the
Taylor series, $\alpha_k=1/k!$, and we recover formula (\ref{40-1}).

\medskip

\emph{Example 2.} Consider the function $g(z_1, \ldots,
z_m)=\prod_{j=1}^m\prod_{k=1}^n 1/(1-t_kz_j)$, as on the left-hand
side of the Cauchy identity (\ref{cauchyid1}). Then
\[
h(z)=\prod_{k=1}^n \frac{1}{1-t_kz}=\sum_{r=0}^{\infty} h_r(t_1,
\ldots, t_n) z^r,
\]
where the $h_r$'s are the complete symmetric functions. Thus
$\alpha_r=h_r(t_1,\ldots, t_n)$ and $ c_{\lambda} = \det
(h_{\lambda_k-k+j})$. In view of the Jacobi-Trudi identity
\[
s_{\lambda}=\det (h_{\lambda_k-k+j})= \det
(e_{\lambda^{\prime}_k-k+j}),
\]
we conclude that $c_{\lambda}=s_{\lambda}(t_1,\ldots,t_n)$, thus
recovering the Cauchy identity (\ref{cauchyid1}).

\medskip

\emph{Example 3.} Consider the function $g(z_1, \ldots,
z_m)=\prod_{j=1}^m\prod_{k=1}^n (1+t_kz_j)$, as on the left-hand
side of the dual Cauchy identity (\ref{cauchyid2}). Then
\[
h(z)=\prod_{k=1}^n (1+t_kz)=\sum_{r=0}^{n} e_r(t_1, \ldots, t_n)
z^r,
\]
where the $e_r$'s are the elementary symmetric functions. Thus
$\alpha_r=e_r(t_1,\ldots, t_n)$, and $ c_{\lambda} = \det
(e_{\lambda_k-k+j})$. The Jacobi-Trudi identity implies that
$c_{\lambda}=s_{\lambda^{\prime}}(t_1,\ldots,t_n)$, and we recover
the the dual Cauchy identity (\ref{cauchyid2}).

\medskip

\emph{Example 4.} Consider the function $g(z_1, \ldots,
z_m)=\prod_{j=1}^m 1/(1-z_j)^{a}$, $a\ge 0$. Then
\begin{equation}\label{add4}
h(z)=\frac{1}{(1-z)^a}=\sum_{r=0}^{\infty} \gamma_r(a)\, z^r, \qquad
  \gamma_r(a)= \frac{\Gamma (a+r)}{\Gamma (a)r!}.
\end{equation}
Thus $\alpha_r=\gamma_r(a)$ in (\ref{a5}) and the coefficients in
the Schur function expansion (\ref{a1}) of $g(z_1,\ldots, z_m)$ are
given by
\begin{equation}\label{a3}
c_{\lambda}=\det (\gamma_{\lambda_k-k+j}(a))_{j,k=1}^m.
\end{equation}
Note that one should assume $|z_j|<1$ to ensure convergence in
(\ref{add4}). Since (\ref{a2}) was obtained by integration over
$|z_j|=1$ there arises the question whether one can still use this
formula. The answer is positive. For, in view of the homogeneity of
the Schur functions
\[
s_{\lambda} (tz_1, \ldots, tz_m)=t^{|\lambda|} s_{\lambda} (z_1,
\ldots, z_m), \qquad |\lambda|=\sum_j \lambda_j,
\]
one can always rescale $z_j \to tz_j$ thus extending the domain for
$z_j$ to include the unit circle.

The determinant in (\ref{a3}) can be evaluated in terms of
$\lambda_k$. Introducing the notation $p_k=\lambda_k-k$, we have
$c_{\lambda}=\det (\gamma_{p_k+j}(a))$. By making use of the
identity
\[
\gamma_{r+1}(a)-\gamma_r(a)=\gamma_{r+1}(a-1)
\]
and elementary operations on the columns of the determinant,
\begin{eqnarray*}
\det (\gamma_{p_k+j}(a))&=& |\gamma_{p_k+1}(a), \gamma_{p_k+2}(a),
\ldots, \gamma_{p_k+m}(a)| \\
& =& |\gamma_{p_k+1}(a), \gamma_{p_k+2}(a-1), \ldots,
\gamma_{p_k+m}(a-1)|\\
&=& \ldots \\
&=& |\gamma_{p_k+1}(a), \gamma_{p_k+2}(a-1), \ldots,
\gamma_{p_k+m}(a-m+1)| = \det (\gamma_{p_k+j}(a-j+1)).
\end{eqnarray*}
Therefore,
\[
c_{\lambda}=\det \left(\frac{\Gamma
(a+p_k+1)}{\Gamma(a-j+1)(p_k+j)!}\right)_{j,k=1}^m=\left\{\prod_{j=1}^m
\frac{\Gamma (a+p_j+1)}{\Gamma (a-j+1)}\right\}  \det
\left(\frac{1}{(p_k+j)!}\right)_{j,k=1}^m.
\]
The determinant on the right-hand side can now be easily evaluated,
\[
\det \left(\frac{1}{(p_k+j)!}\right)_{j,k=1}^m =
\left\{\prod_{j=1}^m \frac{1}{(p_j+m)!}\right\}\prod_{1\le j<k\le m}
(p_j-p_k).
\]
Recalling that $p_j=\lambda_j-j$, we arrive at
\begin{eqnarray*}
c_{\lambda}&=& \left\{\prod_{j=1}^m \frac{\Gamma
(a+\lambda_j-j+1)}{\Gamma
(a-j+1)(m+\lambda_j-j)!}\right\}\prod_{1\le j<k\le m}
(\lambda_j-j-\lambda_j+k)\\ &=& s_{\lambda}(1_m) \prod_{j=1}^m
\frac{\Gamma (a+\lambda_j-j+1) (m-j)!}{\Gamma (a-j+1)
(m+\lambda_j-j)!},
\end{eqnarray*}
where the second equality follows from Weyl's dimension formula
(\ref{eq2:8}). Thus finally (cf. (\ref{cauchyid1}))
\begin{equation}\label{a4}
\prod_{j=1}^m \frac{1}{(1-z_j)^{a}}=\sum_{\lambda} \beta_{\lambda}
s_{\lambda}(1_m) s_{\lambda} (z_1, \ldots, z_m), \quad
\beta_{\lambda}=\prod_{j=1}^m \frac{\Gamma (a+\lambda_j-j+1)
(m-j)!}{\Gamma (a-j+1) (m+\lambda_j-j)!}.
\end{equation}
This identity folds for $a\ge 0$ and $|z_j|<1$. It appears in Hua's
book (Theorem 1.2.5) where it is derived by a different method.

\medskip

\emph{Example 5.} Consider the function $g(z_1, \ldots,
z_m)=\prod_{j=1}^m (1+z_j)^{a}$, $a\ge 0$. Then
\begin{equation}\label{add4a}
h(z)=(1+z)^a=\sum_{r=0}^{\infty} \gamma_r(a)\, z^r, \qquad
  \gamma_r(a)= \frac{\Gamma (a+1)}{\Gamma (a-r+1)r!}
\end{equation}
and the coefficients in the Schur function expansion (\ref{a1}) of
$g(z_1,\ldots, z_m)$ are given by $ c_{\lambda}=\det
(\gamma_{\lambda_k-k+j}(a))_{j,k=1}^m$, with $\gamma_r(a)$ as
defined in (\ref{add4a}). As in the previous example, this
determinant can be evaluated in terms of $\lambda_k$. Now for
$\gamma_r(a)$ we have
\[
\gamma_{r+1}(a)+\gamma_r(a)=\gamma_{r+1}(a+1).
\]
By making use of this identity and elementary operations on columns,
\[
\det (\gamma_{\lambda_k-k+j}(a))_{j,k=1}^m= \det
(\gamma_{\lambda_k-k+m}(a+m-j))_{j,k=1}^m.
\]
On substituting the expression in (\ref{add4a}) for $\gamma_r(a)$
in the determinant on the right-hand side,
\[
\det (\gamma_{\lambda_k-k+m}(a+m-j))_{j,k=1}^m=\left\{ \prod_{j=1}^m
\frac{\Gamma (a+m-j+1)}{(\lambda_j-j+m)!} \right\}\det
\left(\frac{1}{\Gamma (a-j+1-\lambda_k+k)} \right)_{j,k=1}^m.
\]
The determinant on the right can be reduced to the Vandermonde
determinant by elementary operations on columns,
\[
\det \left(\frac{1}{\Gamma (q_k-j)} \right)_{j,k=1}^m=
\left\{\prod_{j=1}^m \frac{1}{\Gamma (q_j-1)}\right\} \prod_{1\le
j<k\le m} (q_k-q_j),
\]
and we arrive at
\begin{eqnarray*}
c_{\lambda}&=& \left\{\prod_{j=1}^m \frac{\Gamma (a+m-j+1)}{\Gamma
(a-\lambda_j+j)(m+\lambda_j-j)!}\right\}\prod_{1\le j<k\le m}
(\lambda_j-j-\lambda_j+k)\\ &=& s_{\lambda}(1_m) \prod_{j=1}^m
\frac{\Gamma (a+m-j+1) (m-j)!}{\Gamma (a-\lambda_j+j)
(m+\lambda_j-j)!}.
\end{eqnarray*}
Thus finally
\begin{equation}\label{a4a}
\prod_{j=1}^m (1+z_j)^{a}=\sum_{\lambda} \beta_{\lambda}
s_{\lambda}(1_m) s_{\lambda} (z_1, \ldots, z_m), \quad
\beta_{\lambda}=\prod_{j=1}^m \frac{\Gamma (a+m-j+1) (m-j)!}{\Gamma
(a-\lambda_j+j) (m+\lambda_j-j)!}.
\end{equation}
This identity, a companion to (\ref{a4}),  holds for $a\ge 0$ and
$|z_j|<1$. If $a$ is a positive integer, say $a=N$, then $1/\Gamma
(N-\lambda_j+j)=0$ for any partition $\lambda$ such that
$\lambda_1\ge N+1$ so that the sum in (\ref{a4a}) is finite. By a
direct computation from the Jacobi-Trudi identity,
\[
c_{\lambda}=\left\{\prod_{j=1}^m \frac{(N+m-j)!)}{(m+\lambda_j-j)!
(N+j-1-\lambda_j)!}\right\} \prod_{1\le j< k \le m}
(\lambda_j-j-\lambda_k+k) = s_{\lambda^{\prime}} (1_N)
\]
and we recover  a particular case of the dual Cauchy identity (cf.
(\ref{cauchyid2}))
\[
\prod_{j=1}^m (1+z_j)^{N}=\sum_{\lambda} s_{\lambda^{\prime}}(1_N)
s_{\lambda} (z_1, \ldots, z_m).
\]


\begin{thebibliography}{99}

\bibitem{AltSim}  Altland A, Simons B D 1999
Field theory of the random flux model {Nucl Phys B} {\bf 562}
No.3: 445-476

\bibitem{AltZirn} Altland A and Zirnbauer M R 1996 Field theory of the quantum kicked rotor
{ \it Phys Rev Lett} {\bf 77}, 4536--4539

\bibitem{AltGnutz} Gnutzmann S and Altland A,
Spectral correlations of individual quantum graphs 2005 {\it Phys
Rev E} {\bf  72} No.5: Art. No. 056215

\bibitem{Balantekin} Balantekin A B 2000 Character expansions, Itzykson-Zuber
integrals, and the QCD partition function  {\it Phys Rev} D(3)
{\bf 62} 085017.


\bibitem{B1} Berezin F A 1974 Quantization {\it Izv Akad Nauk SSSR, Ser Math}  {\bf 38}
1116--1174. English translation: {\it Math USSR-Izv} {\bf 38
(1974)}, no. 5, 1109--1165 (1975)

\bibitem{B2} Berezin F A 1975 Quantization in complex symmetric
spaces {\it Izv Akad Nauk SSSR, Ser Math}  {\bf 39}  363--402;
English translation: {\it Math USSR-Izv} {\bf  9 (1975)}, no. 2,
341--379 (1976)

\bibitem{Budczies} Budczies J, Nonnenmacher S, Shnir Y, and
Zirnbauer M R  2002, (1+1)-dimensional baryons from the SU(N)
color-flavor transformation {\it Nucl Phys} {\bf B 635}, No. 1-2,
309--356.


\bibitem{Bump} Bump D {\it Lee groups}, Springer Science, New York 2004

\bibitem{BG} Bump D and Gamburd A 2006 On the averages of
characteristic polynomials from classical groups. {\it Commun Math
Phys} {\bf 265} 227--274.

\bibitem{ZirnRMT} Conrey J B, Farmer D W and Zirnbauer M R 2005
Howe pairs, supersymmetry, and ratios of random characteristic
polynomials for the unitary group $U(N)$, {\it E-preprint
ArXiv:math-ph/0511024}.

\bibitem{For1} Forrester P J 2006 Quantum conductance problems and the Jacobi
ensemble {\it J Phys A: Math Gen} {\bf 39} 6861 -- 6870.


\bibitem{Forbook} Forrester P J 2005 Log-gases and random
matrices, www.ms.unimelb.edu.au/~matpjf/matpjf.html

\bibitem{FSom} Fyodorov Y V and Sommers H-J 2003 Random matrices close to Hermitian
or unitary: overview of methods and results {\it J Phys A: Math
Gen} {\bf 36} 3303 -- 3347.

\bibitem{F1} Fyodorov Y V 2002 Negative moments of characteristic polynomials of
random matrices: Ingham-Siegel integral as an alternative to
Hubbard-Stratonovich transformation {\it Nucl Phys }   {\bf B621}
643 -- 674.

\bibitem{F2} Fyodorov Y V 2005 On Hubbard-Stratonovich Transformations over
Hyperbolic Domains. {\it J Phys: Cond Matter} {\bf 17} S1915 --
S1928.


\bibitem{FK} Fyodorov Y V and Khoruzhenko B A 2006 On absolute
moments of characteristic polynomials of a certain class of
complex random matrices. {\it E-preprint} math-ph/0602032.

\bibitem{FS} Fyodorov Y V and Strahov E 2002  On correlation functions
of characteristic polynomials for chiral Gaussian unitary
ensemble. {\it Nucl Phys}   {\bf B647} 581--597.

\bibitem{GR} Gradshtein I S and Ryzhik I M  {\it Table of
Integrals, Series, and Products}, 5th ed., A. Jeffrey, editor.
Academic Press, 1994

\bibitem{HO} Harnad J., Orlov, A. Yu. 2006  Fermionic construction of partition
functions for two-matrix models and perturbative Schur function
expansions.  {\it J. Phys A: Math and Gen}  {\bf 39} 8783--8809.

\bibitem{Howe} Howe R 1989 Remarks on classical invariant theory.
{\it Trans Amer Math Soc} {\bf 313} 539–-570.

\bibitem{Hua} Hua L K {\it Harmonic Analysis of Functions of Several
Complex Variables in the Classical Domains}. AMS, Providence,
Rhode Island, 1963

\bibitem{James} James A T 1954 Normal multivariate analysis and
the orthogonal group {\it Annals Math Stat} {\bf 25} 40 -- 75.

\bibitem{JN} Janik R A and Nowak M A 2003 Wishart and anti-Wishart
random matrices  {\it J Phys A: Math Gen} {\bf 36} 3629 -- 3637.

\bibitem{Kad} Kadell K W J 1997 The Selberg-Jack symmetric
functions {\it  Adv Math} {\bf 130} 33 -- 102.

\bibitem{Ka} Kaneko J 1993 Selberg integrals and hypergeometric
functions associated with Jack polynomials. {\it SIAM J Math Anal}
{\bf 24} 1086--1110.

\bibitem{Kaz} Kazakov V A, Staudacher M and Winter T 1996
Exact solution of discrete two-dimensional $R^2$ gravity. {\it
Nucl Phys} {\bf B471} 309 -- 393.

\bibitem{Macdonald} Macdonald I G {\it Symmetric Functions and Hall
Polynomials.} 2nd ed. Clarendon Press, Oxford University Press,
New York, 1995

\bibitem{Mathai} Mathai A M {\it Jacobians of Matrix Transformations and Functions
of Matrix Argument.} World Scientific, Singapore, 1997.

\bibitem{Muirhead} Muirhead R J {\it Aspects of Multivariate Statistical
Theory}, John Wiley \& Sons, New York, 1982.

\bibitem{Ner1} Neretin Y A 2002 Hua-type integrals over unitary groups and
over projective limits of unitary groups {\it Duke Math J} {\bf
114} 239 -- 266.

\bibitem{O} Orlov A Yu 2004 New solvable matrix integrals.
Proceedings of 6th International Workshop on Conformal Field
Theory and Integrable Models. {\it Internat J Modern Phys} {\bf A
19},  May, suppl., 276--293.

\bibitem{OS} Orlov A Yu and Shiota T 2005 Schur function expansion
for normal matrix model and associated discrete matrix models {\it
Physics Lett} {\bf A343 (5)} 384--396.


\bibitem{Perelomov} Perelomov A M  {\it Generalized coherent states
and their applications}, Springer-Verlag, Berlin 1986

\bibitem{SW3} Schlittgen B and Wettig T 2002 Color-Flavor Transformation for
the Special Unitary Group {\it Nucl Phys} {\bf B 632} 155 -- 172.


\bibitem{SW1} Schlittgen B and  Wettig T 2003 The Colour-Flavour
transformation and Lattice QCD   {\it Nucl Phys B-Proc Suppl} {\bf
119}, 956--961

\bibitem{SW} Schlittgen B and  Wettig T 2003 Generalizations of some integrals
over the unitary group. {\it J Phys A: Math and Gen} {\bf 36} 3195
-- 3202.


\bibitem{SM} Simon S H and Moustakas A L 2006 Crossover from
conserving to lossy transport in circular random matrix ensembles
{\it Phys Rev Lett} {\bf  96}  No. 13: Art. No. 136805.

\bibitem{Stanley} Stanley R P 1989 Some combinatorial properties of Jack symmetric
functions {\it  Adv Math} {\bf 77}  76 -- 115.


\bibitem{WW} Wei Y and Wettig T 2005 Bosonic color-flavor
transformation for the special unitary group. {\it J Math Phys}
{\bf 46} Art. No. 072306.


\bibitem{Yan} Yan Z M 1992 A class of generalized hypergeometric functions in
several variables.  {\it Canad J Math} {\bf 44}  1317--1338.

\bibitem{Z1} Zirnbauer M R 1996 Supersymmetry for systems with
unitary disorder: Circular ensembles {\it J Phys A: Math Gen} {\bf
29} 7113 -- 7136.

\bibitem{Z2} Zirnbauer M R 1999 The Color-Flavor Transformation and a new
approach to quantum chaotic maps. {\it Proceedings of the XIIth
International Congress of Mathematical Physics, Brisbane 13-19 July
1997}, ed. D De Wit, A J Braken, M D Gould, P A. Pearce
(International Press Inc. Cambridge, MA) pp. 290 -- 297.

\bibitem{Zirnhall}  Zirnbauer M R 1997
Toward a theory of the integer quantum Hall transition: Continuum
limit of the Chalker-Coddington model {\it J Math Phys} {\bf 38}
No. 4, 2007--2036


\bibitem{Zirntrento} Zirnbauer M R 2006, Howe Duality and the Color-Flavor transformation,
talk at: New Directions in Nonperturbative QCD, Trento 27-31
March, 2006, available at $
http://www.ect.it/Meetings/ConfsWksAndCollMeetings/ConfWksDocument/2006/Talks/27-31_March/zirnbauer.pdf$


\bibitem{ZS} \.Zyczkowski K and Sommers H-J 2002 Truncations of random
unitary matrices. {\it J. Phys. A: Math. Gen.} {\bf 33} 2045 --
2057.



\end{thebibliography}
\end{document}